## PERSPECTIVE

# Current Density in Solar Fuel Technologies

Valentino Romano,[a,*] Giovanna D'Angelo,[a] Siglinda Perathoner[b] and Gabriele Centi[b,*]

Solar-to-fuel direct conversion devices are a key component to realize a full transition to a renewable-energy based chemistry and energy, but their limits and possibilities are still under large debate. In this perspective article, we focus on the current density as a fundamental figure of merit to analyse these aspects and to compare different device configurations and types of solar fuels produced from small molecules such as $H_2O$, $CO_2$ and $N_2$. Devices with physical separation of the anodic and cathodic zones, photoelectrochemical-type (PEC) or with a photovoltaic element integrated in an electrochemical cell (PV/EC), are analysed. The physico-chemical mechanisms involved in device operation that affect the current density and relations with device architecture are first discussed. Aspects relevant to device design in relation to practical use are also commented. Then discussion is moved towards the relevance of these aspects to compare the behaviour in the state-of-the-art of the conversion of these small molecules, with focus on solar fuels from $CO_2$ and $N_2$ conversion, analysing the gaps and perspectives. The still significant lack of crucial data, notwithstanding the extensive literature on the topic, has to be remarked, particularly in terms of the need to operate these cells in conjunction with sun concentration (in the 50-100 SUN range) which emerges as the necessary direction from this analysis, with consequent aspects in terms of cell and materials design to operate in these conditions. The work provides a guide for the optimisation of the investigated technology and the fixing of their practical limits for large-scale applications.

## Introduction

The transition towards renewable energy sources offers the possibility of solving the environmental burdens due to increasing greenhouse gas emissions, but it also represents a clear economic and innovation opportunity.[1] In this context, the use of solar fuels (SFs), *i.e.*, chemicals produced through reactions triggered directly or indirectly by sunlight, is pivotal for the widespread use of renewable energy (above ~50-60%)[2] because they play an essential role in (i) mitigating renewable energy fluctuations, (ii) storing energy on seasonal/yearly bases and (iii) transporting energy to long distances, allowing the implementation of a world-scale renewable-energy economy. They will serve for a smoother and progressive substitution of fossil fuels as primary energy source,[3–8] since SFs use in current energy infrastructure can take place with minimal adaptation.

The production of SFs is discussed in reviews,[9–12] mainly focused on three classes of SFs realised by photo- or photo-electrocatalytic processes: (i) $H_2$ from water, (ii) fuels and chemicals by $CO_2$ reduction (CO, HCOOH, $CH_3OH$, C2+ alcohols, hydrocarbons and acids, where C2+ indicates products with two or more carbon atoms), and (iii) ammonia by reduction of $N_2$. Reaction (i) is common to the 2nd and 3rd reactions to provide the hydrogen for the reduction of $CO_2$ or $N_2$ in an integrated process. More correctly, these reactions do not involve the intermediate formation of molecular $H_2$, but rather the direct use of the hydrogen-equivalent species (H+/e-) deriving from

a. Dept. MIFT, University of Messina, Vl.e F. Stagno D'Alcontres 31, 98166 Messina, Italy.
b. Dept. ChiBioFarAm and ERIC aisbl, University of Messina, Vl.e F. Stagno D'Alcontres 31, 98166 Messina, Italy.

water (or other molecules) anodic oxidation. Here we limit the discussion only to the case of direct solar utilization, *i.e.*, artificial photosynthesis devices. Therefore, the cases of electrolysis or co-electrolysis processes are not included, as well as thermochemical solar processes. In addition, the discussion is limited to the production of $H_2$, $NH_3$ and $CO_2$ hydrogenation products, excluding other cases (such as $H_2O_2$) which, although interesting, may defocus discussion.

The most studied reaction in SFs research is the conversion of $H_2O$ into $H_2$ because (i) $H_2O$ is an abundant resource; (ii) the combustion reaction of $H_2$ releases only water ($H_2+\frac{1}{2}O_2 \rightarrow H_2O$); (iii) $H_2$ has a gravimetric energy (141.9 MJ·kg⁻¹) which is far higher than that of typical fossil fuels like gasoline (47.5 MJ·kg⁻¹) and diesel (44.8 MJ·kg⁻¹).[9,13,14] Note that the energy density by volume of $H_2$ shows the opposite trend, being the main limit from the application perspective. In fact, even when compressed to 700 bar, the volumetric energy density of $H_2$ is 5.6 MJ·L⁻¹ with respect to 32.0 MJ·L⁻¹ for gasoline. For all these reasons, there is a large recent push in Europe and worldwide in establishing a strategy to produce green $H_2$.[15]

However, the limits for an $H_2$ economy are related to its storage and distribution.[16–20] For these reasons, SFs produced from $CO_2$ or $N_2$ will also play a crucial role. For example, products of $CO_2$ reduction reaction ($CO_2RR$) could be used directly as energy vectors substituting those from fossil fuels (without or with minimal changes in currently used devices). Furthermore, SFs from $CO_2RR$ have good gravimetric energy (55.5 MJ·kg⁻¹ for methane, $CH_4$, and 20 MJ·kg⁻¹ for methanol, $CH_3OH$),[9,21] and acceptable volumetric energy (15.6 MJ·kg⁻¹ for $CH_3OH$, while 0.04 MJ·kg⁻¹ for $CH_4$ at standard conditions) but they have the disadvantage of needing to recapture the emitted $CO_2$ to make their use completely sustainable. As concerns $NH_3$, since it can be liquefied under mild pressure, it can be used as both hydrogen and energy vector,[22] the latter in fuel cells and ammonia turbines. The main advantage is that the combustion product is $N_2$ which can be released to the atmosphere. The associated gravimetric and volumetric energies for liquid $NH_3$ are 18.8 $MJ \cdot kg^{-1}$ and 11.5 $MJ \cdot L^{-1}$, respectively, which are comparable to those of $CH_3OH$. With respect to formic acid, HCOOH another $H_2$ vector obtained from $CO_2$ reduction,[22] $NH_3$ presents several advantages: (i) the hydrogen content by weight is higher in $NH_3$ (~18%) with respect to HCOOH (4%); (ii) $NH_3$ decomposes producing $N_2$ ($NH_3 \rightarrow 1/2\ N_2 + 3/2\ H_2$), while HCOOH decomposition releases $CO_2$ (HCOOH $\rightarrow CO_2 + H_2$); (iii) the production and transport of liquid $NH_3$ are already implemented large-scale processes, so such infrastructures can be readily used for distributing $NH_3$. Thus, $NH_3$ is an attractive carbon-free energy or hydrogen vector, which can be implemented as SFs by using sunlight to trigger $N_2$ reduction reaction (NRR).[23]

Solar-to-fuel conversion is thus a complementary crucial technology to green $H_2$ produced by electrolysis to move toward a carbon-neutral future. In this review only integrated systems which combine in a single device the functions of harvesting sunlight and its use to produce SFs will be discussed.[24–33] From the first report by Fujishima and Honda in 1972,[34] several architectures have been proposed for the efficient synthesis of SFs.[28,33] We can lump the proposed architectures in three main classes: (i) particulate photocatalysis (PP), (ii) photoelectrochemical (PEC) devices and iii) photovoltaic-driven-electrocatalysis (PV/EC). In PP, nanoparticles of semiconductors are suspended in the electrolyte and conduct both redox reactions on their surfaces. Note that the term PP is not commonly used, but we believe it is useful to differentiate the photocatalysis approach from those such as PEC or PV/EC where there is physical separation between the anodic and cathodic processes. In PEC systems reduction and oxidation take place at two electrodes (a photoelectrode and a counter electrode or two photo-electrodes). In PV/EC devices a solar cell is connected to an electrolyser which produces the SF. Focus in this review is on these devices having formation of anodic and cathodic products in separate streams, because those actually more promising to produce solar fuels, as commented later.

These devices are complex with many aspects determining their behaviour. It is thus necessary to use figures of merit (FoM) for their direct and proper comparison.[35,36] Systems efficiencies are usually employed as FoMs, which are defined as the ratio of the total output power (in the form of electricity and/or energy stored in the SFs) to the total input power (i.e., electricity and/or solar power).[35] For the case of $H_2$ production, the solar-to-hydrogen conversion efficiency ($\eta_{STH}$) is often employed as FoM, which is defined as:

$$\eta_{STH} = \left| \frac{|J_{SC}| \cdot 1.23\ V \cdot \eta_F}{P_{in}} \right|_{AM1.5G} \qquad (1)$$

where $J_{SC}$ is the short circuit current density, $\eta_F$ is the Faradaic efficiency for $H_2$ evolution and $P_{in}$ is the power of the incident radiation (which is measured at AM1.5G condition).[37,38] Solar-to-fuel (STF) efficiency ($\eta_{STF}$) is the equivalent definition applied to products different than hydrogen. These FoMs are defined for the whole device, so they do not provide information about the occurring physico-chemical processes and how to improve the resulting performances. Several other FoMs have been introduced to characterise half-cell performances, such as the ideal regenerative cell efficiency, the ratiometric power-saved and the applied bias photon-to-current component metric.[35]

Usually, calculating these FoMs is not trivial, since corrections are needed to account for many phenomena (such as mass-transport, resistance due to the device geometry, etc.) and the choice of the counter electrode/device assembly may lead to significant variations in the computed results.[35] This can make comparisons of results from different research groups challenging, especially if all the experimental conditions and calculations are not explicitly reported. A fair comparison with already existing SF generation systems can become cumbersome, as different working principles can be involved.

In this perspective, we propose current density ($J$) as a FoM to analyse performances and compare SFs approaches and devices. This choice is due to several reasons: (i) a detailed analysis of $J$ can provide many information about the physico-chemical processes occurring during the device operation; (ii) it is readily obtained from simple electrical or electrochemical characterizations (thus, it is easily accessible) and (iii) it represents a useful FoM for comparisons with already existing technologies (for example, commercial electrolysers work with $J$ ~0.5-2 A $cm^{-2}$).[39,40] Furthermore, from eq. 1 it is clear that by fixing the operating voltage and maximising $\eta_F$ (ideally 100%), $\eta_{STF}$ is a function of $J$ only. As will emerge later from the discussion, focusing attention on $J$ as FoM allows to bring out the need to operate devices in conjunction with sun concentration, an aspect crucial for practical implementation, but essentially not investigated and that has also significant implications in terms of materials and cells needed to operate under this mode. Moreover, $J$ is also an effective FoM from an industrial perspective, because it provides a direct indication on the productivity and other cell design characteristics.

To properly discuss the use of $J$ and relevance in comparing devices for the different indicated SFs, this paper is organised as follows: the first part deals with the basic principles behind SF generation and the impact of device architecture. For convenience we divide the processes involved during SF generation in four steps, similarly to other literature studies.[41,42] All these aspects influence the value of $J$, so in the following parts we discuss the physics and chemistry behind the optimization of such processes and emphasise the information that can be obtained by a careful analysis of $J$. The last part is dedicated to analysing the different design configurations of PEC cells, their use in relation to the different SFs that can be produced and their potential as high-$J$ generating systems.

Note that the aim is not to make a systematic review and discussion of the state-of-the-art, but to offer concepts and clues to rethink the topic from a different viewpoint. Only selected references have been considered.

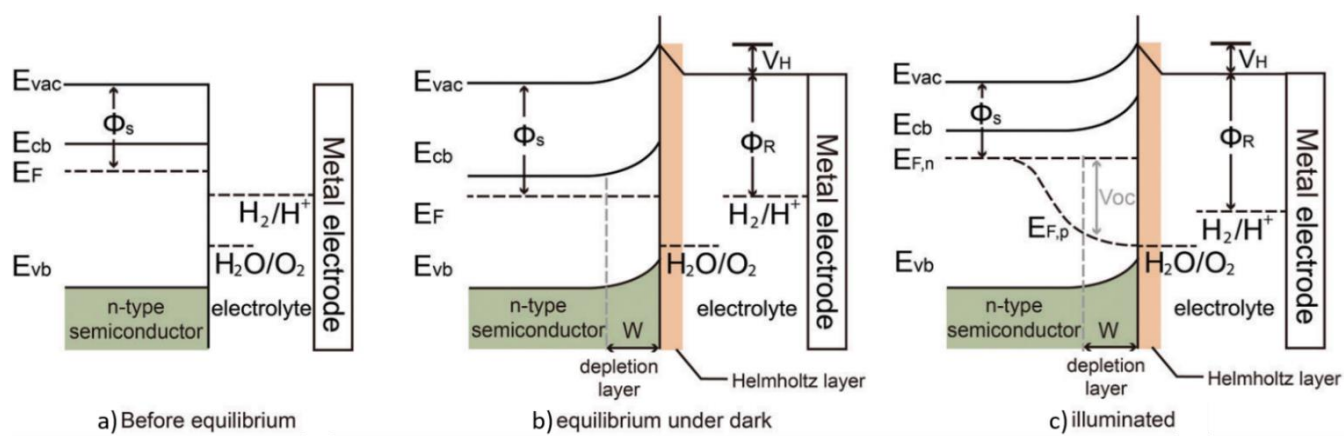

**Figure 1.** Basic photoelectrochemistry of PEC devices a) before equilibrium, b) in dark conditions and c) under illumination. $E_{vac}$: vacuum energy; $E_{cb}$: conduction band; $E_{vb}$: valence band; $\Phi_S$: work function of the semiconductor making up the photoelectrode; $\Phi_R$: electrolyte work function; $V_H$: potential drop due to the Helmholtz layer; $E_{F,n}$: electron quasi-Fermi level; $E_{F,p}$: hole quasi-Fermi level. Figure taken from ref.[43] Copyright Royal Society of Chemistry @2017.

## Solar to fuel conversion

### Working principles

The synthesis of SFs involves a complex machinery of events which, in a simplified approach, can be divided in:

a) Photon absorption and generation of free charge carriers
b) Separation and transport of charge carriers
c) Catalytic conversion
d) Mass transfer

In PEC systems, all these steps are performed within the PEC cell, whereas in PV/EC design charge carriers are photo-produced and separated in the PV component, while catalytic conversion and mass transfer take place in the EC unit.

Herein, we consider a PEC system comprising a photoanode, an electrolyte and a metallic counter electrode (Figure 1a). When these three components are connected together, a charge flow is triggered that brings the whole system to equilibrium (*i.e.*, to the condition where the chemical potential is constant throughout the device).[43,44] For the case of a *n*-type semiconducting photoanode, the equilibrium at dark conditions is reached through an electron flow from the semiconductor to the solution (Figure 1b).[43,44] This causes the formation of a depletion layer inside the photoanode and an Helmoltz layer in the electrolyte.

The region depleted from electrons is characterised by the presence of fixed positive ions, leading to the formation of a built-in electric field which induces the "up" bending of the electronic bands. The same phenomenon occurs for *p*-type semiconducting photocathodes, with the difference that a "down" band bending occurs. In both *n*- and *p*-type cases, the band bending induces an electric field which opposes the motion of electrons (for *n*-type) or holes (for *p*-type) from the photoelectrodes to the electrolyte once equilibrium of the chemical potential is reached (*i.e.*, when the Fermi level of the semiconductor equals the chemical potential of the electrolyte redox couple).

Upon illumination, the light-harvester of the photoelectrode absorbs photons whose energy equals or exceeds that of its band gap ($E_g$). When this happens, electron-hole pairs are generated which (depending on the properties of the semiconductor, as described in the following sections) can form a stable bound state (exciton) or dissociate into free charge carriers. Such non-equilibrium density of electrons and holes is described through the concept of quasi-Fermi level (one per type of charge carrier, Figure 1c).[43–45] The difference between the values of these quasi-Fermi levels represents the photo-produced voltage, *i.e.*, the photovoltage or open circuit voltage $V_{OC}$.[43,45] It is worth emphasising that photo-generated carriers are responsible to trigger the reactions for SFs production. However, such processes can be activated only if several constraints are met. In particular, splitting reactions are endergonic (*i.e.*, they require energy to take place). For the case of $H_2$ from $H_2O$, $CH_3OH$ and $CH_4$ from $CO_2$, and $NH_3$ from $N_2$ these equations formally describe the splitting processes:

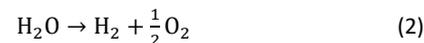
$$H_2O \rightarrow H_2 + \frac{1}{2}O_2 \tag{2}$$

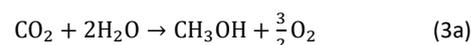
$$CO_2 + 2H_2O \rightarrow CH_3OH + \frac{3}{2}O_2 \tag{3a}$$

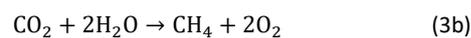
$$CO_2 + 2H_2O \rightarrow CH_4 + 2O_2 \tag{3b}$$

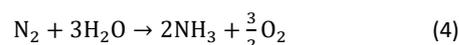
$$N_2 + 3H_2O \rightarrow 2NH_3 + \frac{3}{2}O_2 \tag{4}$$

The overall water splitting (OWS) reported in (eq. 2) has a Gibbs free energy $\Delta G° = 237$ kJ·mol$^{-1}$, the $NH_3$ synthesis (eq. 4) needs $\Delta G°= 680$ kJ·mol$^{-1}$, while the $CO_2$ reduction into $CH_3OH$ (eq. 3a) or $CH_4$ (eq. 3b) require $\Delta G°= 689$ kJ·mol$^{-1}$ and $\Delta G°= 800$ kJ·mol$^{-1}$ respectively (reported values correspond to 25 °C and 1 bar conditions).[21,46,47] The $\Delta G°$ values can be readily converted into potential difference through the equation

$$\Delta E^0 = \frac{\Delta G^o}{nF} \tag{5}$$

where *n* is the number of electrons involved in the reaction and *F* is the Faraday constant (96485.3365 C·mol$^{-1}$).[48] Since OWS, $N_2$ to $NH_3$ and $CO_2$ to $CH_3OH/CH_4$, involve, respectively, 2, 6, 6 and 8 electrons, the minimum potential difference required to drive each reaction are, respectively, 1.23, 1.17, 1.19 and 1.03 V. Consequently, the minimum photon energy needed to trigger these solar-to-fuel reactions are 1.23, 1.17, 1.19 and 1.03 eV.[21,46]

However, these values consider only the thermodynamic limits from the reaction itself. Indeed, the minimum energy required for the practical production of SFs is actually higher, mainly because of ohmic and charge transport losses, kinetic overpotentials and reactor geometry determining cell resistances[49] (for example in the case of OWS values span between 1.8 and 2.4 eV).[10,50] In general, increasing the potential with respect to RHE (reversible hydrogen electrode), favours higher formation rates and thus the electrode productivity. However, in NRR and CO2RR, higher potentials also favour the side reaction of $H^+/e^-$ recombination to form $H_2$. Thus, to maintain high the selectivity to the target product, the overall overpotential is kept low (from 0.1-0.2 V vs RHE,[51] to 0.4-0.5 V) which implies low reaction rates and electrode productivities.

Thus, semiconductors with such band gaps must efficiently produce enough photo-voltage to conduct the desired reactions. The band edge absorption lies at wavelengths higher than 600 nm, which reduces drastically the utilization of sunlight, consisting of IR $(2500 - 700$ nm), visible $(700 - 400$ nm) and UV $(400 - 300$ nm) regions which represent 52%, 43% and 5% of solar radiation.[43]

Once electrons and holes are photo-generated, they must be kept separated and driven towards the interface formed with the electrolyte (in the PV/EC design charges are driven to the PV current collector and then to the EC component). When charges reach the active sites of the surfaces, electrons and holes are used for the redox reactions of cathodic conversion of protons and electrons to $H_2$ (OWS) or reduction of $N_2$ or $CO_2$ (NRR and CO2RR, respectively) while generation of $O_2$ (OWS) or other oxidation reaction at the anode side. For these processes to take place favourably, charge carriers inside the photoelectrode must lie into proper energy levels, *i.e.*, the conduction band minimum (CBM, for electrons) and valence band maximum (VBM, for holes).[21] Thus, electrons can reduce a species if the CBM lies above the reduction potential, while holes can drive oxidation of a reactant if the VBM lies below the oxidation Finally, the circuit is electrically closed by the motion of the ions within the electrolyte. This step is crucial to obtain high reaction rates because it is responsible to bring "new" active species from the electrolyte to the particle surface where they can be used for the redox processes. Thus, a fast mass transfer is fundamental to enhance the kinetics of the reactions and keep high $J$ values.

This sequence of processes occurs in both PP and PEC cases, but in the latter the anodic and cathodic processes occurs in physically separated zones, allowing thus a separation of the anodic and cathodic products of reaction with the many inherent advantages (reduction of the costs of separation and safety, etc.). A measurable photocurrent is present between the anodic and cathodic compartments, and thus current density ($J$) is a direct measure of the process. In PP, instead, no measurable $J$ is present and for this reason no longer discussed here. The PV/EC architecture shares with PEC the compartmentalisation in anodic and cathodic zones, but the process of photogeneration of the current occurs in a separate PV element and is not integrated in one of the cell zones, typically the anodic one. Also for PV/EC the photocurrent can be measured

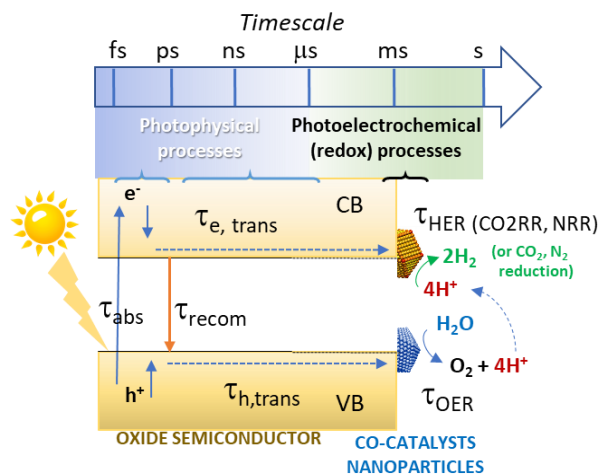

**Figure 2.** Indicative timescale ($\tau$) of the photophysical and (photo)electrochemical processes in water splitting. CB: conduction band; VB: valence band; trans: transfer. Elaborated from ref.[41] Copyright Springer @2015.

and is an indication of the process effectiveness.

This brief discussion shows that SFs production can be highly efficient only if these four steps are wisely optimised (a thorough discussion of the implications of each step is presented in the following sections). There are, however, other aspects that must be considered. A careful engineering of devices for SFs generation is needed. Many architectures are based on heterojunction systems, *i.e.*, different materials are used synergistically with the aim to optimise each process. For example, charge transporting layers are used to selectively drive electrons or holes to one specific direction, interlayers are added to improve charge transport and passivate defects formed at the interfaces between two materials, and electrocatalysts are deposited onto the interface in contact with electrolyte to increase the catalytic activity of the reactions. Thus, the architecture of devices for SFs production shows several additional possible resistance and interfaces limiting the overall $J$, as discussed later. The understanding of these effects is crucial to obtain high performances, but often not specifically investigated.

In addition, it must be remembered that high $J$ implies fast charge carrier transport for each step of device machinery. However, there is a significant difference in the timescales ($\tau$) involved during light harvesting, charge separation and transport, redox reactions, and electrolyte transport. In fact, charge generation and separation are fast processes ($\tau \sim$ fs) while charge transport, redox reactions and mass transfer take place at slower timescales ($\tau \sim \mu$s, $\mu$s/ms and ms/s, respectively) as shown in Figure 2.[41] Thus, when carriers reach the active sites of the reactions, they may "wait" for a long time before being used to trigger reduction or oxidation processes because of slow catalytic activity or sluggish motion of the ions inside the electrolyte. The use of an electrocatalyst can accelerate the redox reaction. However, in most of the cases, its main roles are to reduce the rate of charge recombination and to avoid alternative reactions by different mechanisms (hetero- and phase-junctions, Schottky junction, interface engineering, facets and defects, surface functionalization), rather than effectively catalysing the process of charge transfer.[52–55]

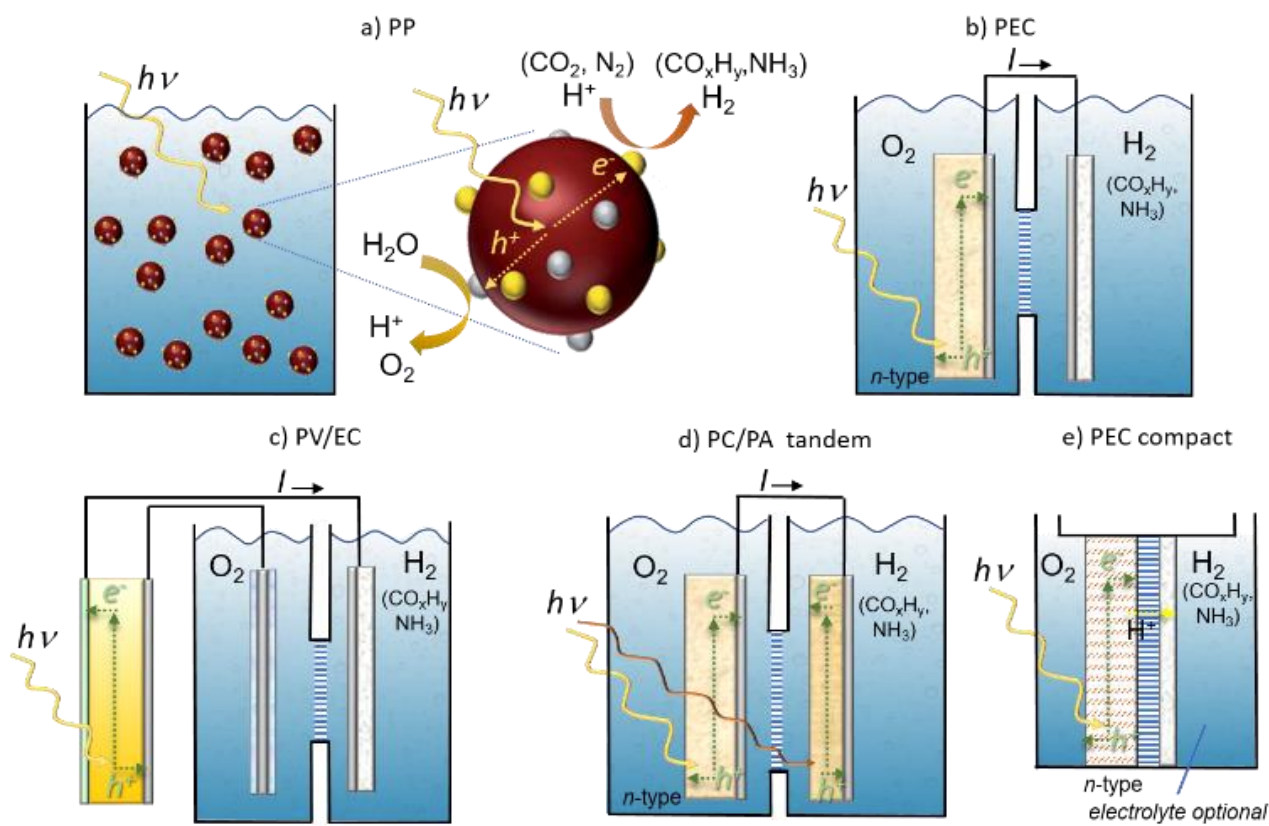

**Figure 3.** Schematic representation of the different architectures used for SFs production. a) particulate photocatalysis (PP); b) photoelectrochemical catalysis (PEC); c) photovoltaic-driven catalysis (PV/EC); d) photocathode/photoanode tandem configuration (PC/PA); e) PEC compact (optional use of the electrolyte). Modified from ref.[58]

**Technologies and approaches**

In the previous sections, the three main approaches used to produce SFs were introduced: PP, PEC and PV/EC. Although the PP case, as commented, is not discussed in detail here because a $J$ cannot be measured and used as FoM, we recall it here briefly as a comparative element to understand better the differences between PEC or PV/EC and PP case, the latter being one of the most investigated, because simpler (even with many limitations). Herein, we thus discuss the pros and cons of these different device configurations for solar-to-fuel conversion.

In PP systems (Figure 3a), nano- or micro-particles (made up by semiconductor materials) are suspended inside a water solution and use light to directly drive both the oxidation and reduction reactions on their surfaces.[52,56,57] Typically, the two redox reactions occur on different crystal facets of the particle modified by adding different co-catalyst nanoparticles. Note that these added nanoparticles (used to promote the redox electron transfer reactions) are typically indicated in literature with the term co-catalyst (rather than only catalyst) because acts as co-elements to enhance the electron transfer capability of the semiconductor substrate itself. For OWS, PP is the simpler approach to produce $H_2$ requiring only semiconductor particles (suspended in an aqueous electrolyte) and light to operate, but many issues limit its large-scale use. First, it is difficult to find a material that can meet the energetics requirements for both hydrogen and oxygen evolution reactions (HER and OER, respectively) while being an optimal light harvester and stable against corrosion.

Using co-catalysts can mitigate some of these constraints and several approaches have been proposed (such as defect engineering, Z-scheme, elevated temperature photocatalysis, etc.)[9,59] with the aim to increase the performances of such devices but so far, the conversion efficiencies are low (for the case of OWS, $\eta_{STH}$ ~1%).[9] In addition, there are technological limitations in the PP approach, notwithstanding the inherent simplicity. The light scattering due to photocatalyst particles suspended in the aqueous solution limits the effectiveness of light penetration to few cm. Moreover, there are additional costs to avoid the deposition of the photocatalyst particles and to separate the $H_2$ and $O_2$ that are evolved in the same container. In addition, there are severe security issues (the $H_2/O_2$ evolving mixture is inside the explosively region) and back/side reactions are relevant, making particularly for $CO_2RR$ and NRR cases unsuited the PP approach.

The key elements of PEC devices are the following (Fig. 3b):
1. *A photoelectrode*: usually this is a semiconductor layer or a bulk-heterojunction component where a light harvester is sandwiched between the selective charge extraction layers. In this way, the resulting built-in electric field favours the separation of electron-hole pairs (generated in the absorber) and drives them to the back contact or to the interface with the electrolyte, depending on the considered redox process. A (co)-catalysts can be deposited at the photoelectrode/electrolyte interface to promote the redox reactions. Protective layers can also be needed to prevent the direct contact between the photoelectrode and the electrolyte (this configuration is usually referred to buried junction) since the former can suffer from corrosion.

2. *A cell system*: it comprises the hosting structure with separate anodic and cathodic zones, a membrane separating these two compartments and wires to transport the electrons from (typically) the anodic to the cathodic zone. Membranes and wires are necessary to close the ionic and electronic circuit, while maintaining the physical separation between the anodic and cathodic zones.

3. *A counter electrode*: in OWS, this could be a simple metallic electrode where HER occurs, but in $CO_2RR$ and NRR cases, the electrode has a more sophisticated multicomponent structure, being necessary to maximize the aimed reaction, while minimizing the side HER reaction.

It is worth noting that since reduction and oxidation reactions take place in different compartments, issues related to separation and recombination of the products, side reactions and safety are avoided, overcoming the intrinsic limitations of the PP case. However, the higher complexity of the resulting devices results in additional steps and sources of resistances, briefly outlines as follow:

1. At the photoelectrode
   a) Photon absorption and generation of free charge carriers
   b) Separation and transport of charge carriers
   c) Catalytic conversion (into the reduced or the oxidised species, depending on the reaction occurring at the photoelectrode)
   d) Evolution of the products in the form of gaseous bubbles (avoiding the stable deposition on the surface of photoelectrode) and inhibition of side and recombination reactions
   e) Collection and transport, through an external wire, of the other charge carrier at the cathode

2. At the membrane
   a) Mass transfer through the membrane
   b) Inhibition of cross-over from the counter electrode zone.

3. At the counter electrode
   a) Mass transfer of active species from the membrane to the catalyst
   b) Transport of the charge carriers (generated by the photoelectrode and incoming through the wire) to the catalyst sites
   c) Evolution of the products in the form of gaseous bubbles (avoiding the stable deposition on the surface of photoelectrode) and inhibition of side and recombination reactions.

This intrinsically complex machinery, where all the steps are highly correlated and not independent (as often assumed), require special care and optimal engineering of the whole device to obtain high efficiencies. As a result, the structure of PEC photoelectrodes is complex, resulting in higher fabrication costs. For the case of OWS, currently PEC systems show 2 % < $\eta_{STH}$ < ~ 19 %,[9] depending on the photoelectrode components.

The third device architecture is that of PV/EC (Figure 3c), where a solar cell is connected in series to an electrolyser: the redox reactions are carried out by the latter, using the charge carriers photo-generated in the PV component. It's worth mentioning that, with this kind of architecture, the solar cell is not in direct contact with the electrolyte, preventing detrimental effects due to corrosion. Furthermore, since both photovoltaic and electrolyser technologies have reached high efficiencies, the PV/EC approach is the most promising solution for sustainable SFs production,[9] with the current record for OWS of $\eta_{STH}$ = 30%,[60] although performances are not stable.

Another promising design, which exploits the possibility for a two-step light absorption somewhat similar to Z-scheme in PP,[61,62] is the tandem PEC configuration.[63,64] In this case, two photoelectrodes are used synergistically to increase both photocurrent and photovoltage, i.e., by maximising the absorbed portion of the solar spectrum and optimising the electronic band gaps/band edges positions.

The main configuration of PEC tandem cells is the photocathode/photoanode (PC/PA), consisting of two photoelectrodes connected with an external wire (Figure 3d). To optimise the light harvesting, one of the photoelectrodes and the separation membrane must be transparent to the light absorbed by the second photoelectrode, a requirement that limits the choice of materials used in such architecture. Thus, although tandem configurations have the potential to deliver high efficiencies, the expensive costs of fabrication often hinder their practical use.

In addition to configurations where the two electrodes are immersed in the electrolyte far from the membrane separating the two hemicells (Figures 3b-d), a compact configuration where the two electrodes are positioned on the two sides of the membrane can be realised (Figure 3e). This "monolithic" configuration minimizes resistance mass transfer and ohmic losses, as described in the following sections.

As a final consideration, note that the efficiency of the systems can be increased by the addition of low-cost Fresnel lenses acting as a sun-concentrator. Depending on the design, sun concentration factors up to three orders of magnitude can be reached,[65] but even a quite simple design can allow to increase light intensity by a factor up to 30-50 at a minimal additional cost. However, this assembly can easily induce temperature rise even above 100°C. Since most of the current cell designs and materials cannot run in a temperature range of around 100-200°C, further research is needed to address the possible applications of (low cost) sun-concentration systems to enhance the PEC performances.

### Reactions

Herein, we focus our attention to the conversion of $H_2O$, $N_2$ and $CO_2$ into SFs. Such reactions present different features, but all are multi-electron reactions. If only the cathodic reaction is considered, HER is a 2-electrons reaction (eq. 6), but a 4-electron reaction by considering full OWS (*e.g.*, the formation of $O_2$). NRR is a 6-electron reaction (eq. 7), while $CO_2RR$ depends on the target product. For example, CO and HCOOH formation (eq.s 8a and 8b) are 2-electrons reactions, methanol production is a 6-electrons reaction (eq. 9), while $CH_4$ formation is a 8-electrons reaction (eq. 10). More complex reactions have been proven to be feasible involving the formation of C-C bonds, such as the 8e⁻ reduction to acetic acid ($CH_3COOH$, eq. 11)[66] or the 18e⁻ reduction to isopropanol ($CH_3CH(OH)CH_3$, eq. 12).[67,68]

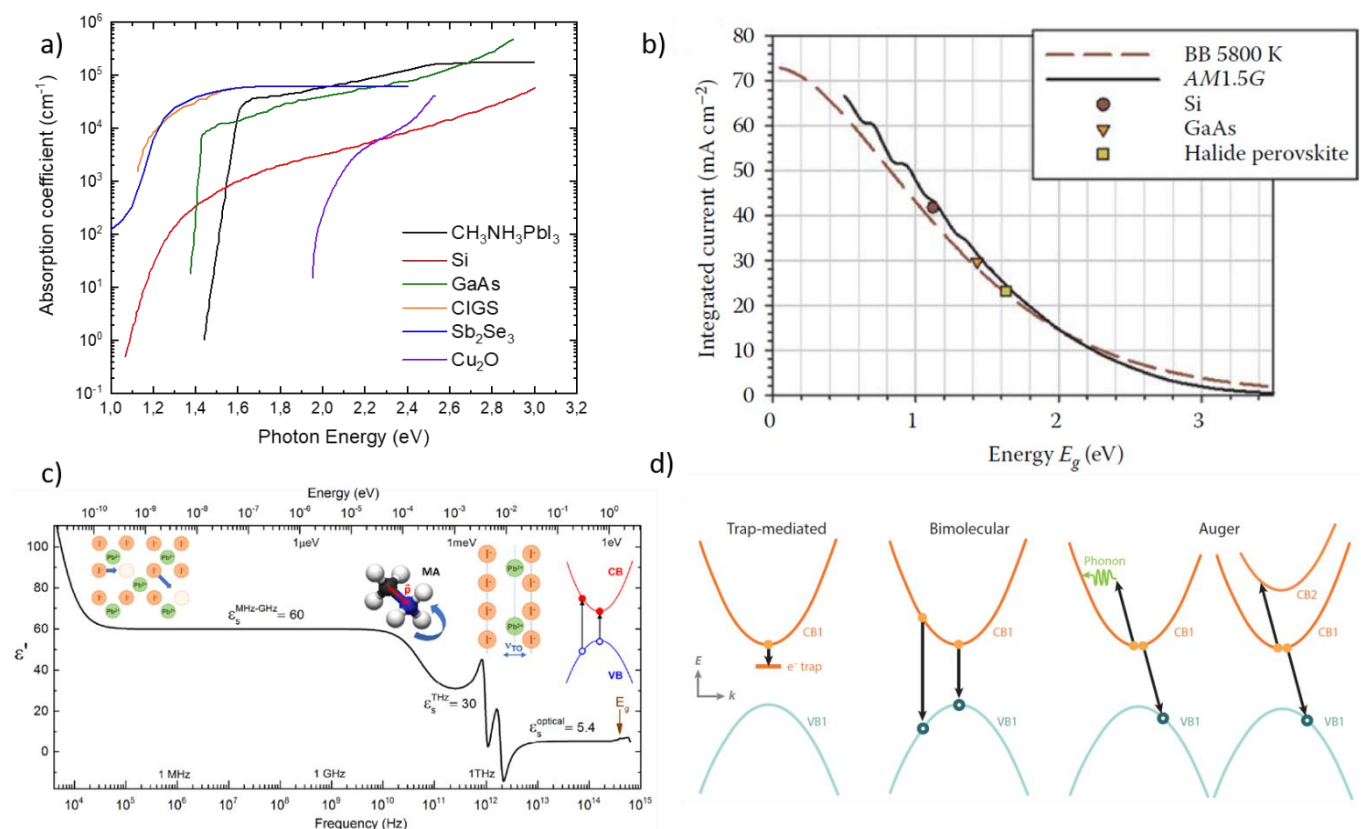

**Figure 4.** a) Absorption coefficient of several photo-active materials for photovoltaic and solar-to-fuel conversion devices (data for $CH_3NH_3PbI_3$, Si and GaAs are taken from ref.[98]; CIGS from ref.[99]; $Sb_2Se_3$ from ref.[100]; $Cu_2O$ from ref.[101]). b) Theoretical integrated current density as a function of the energy bandgap of the light harvester by considering, as photon flux, the radiation emitted by a black body at 5800 K (dashed brown line) and the AM 1.5G SUN spectrum (solid black line). Points indicate recorded values for representative technologies (Si, GaAs and halide perovskites). Data taken from ref.[99] Copyright CRC Press @2018. c) Representation of the frequency dependence of the real part of the dielectric function for the case of $CH_3NH_3PbI_3$. Several phenomena contribute to the overall behaviour of such function: migration of ions ($\leq 100$ kHz); MA⁺ ($CH_3NH_3^+$) reorientations (100 GHz); optical phonon resonances (1-2 THz); electronic interband transitions (400-800 THz). Data taken from ref.[102] Copyright ACS publications @2018. d) Schematic representation (in E-k space) of the main recombination pathways in semiconductors: trap mediated, bimolecular and Auger. The symbols CB and VB stand for conduction and valence band respectively. The numbers 1 and 2 consider the eventual spin-orbit coupling effect arising in some materials (such as metal halide perovskites, for which this figure was proposed). Taken from ref.[103] Copyright Annual Reviews @ 2016.

$$2H^+ + 2e^- \rightarrow H_2 \tag{6}$$

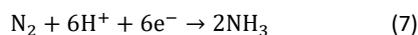

$$N_2 + 6H^+ + 6e^- \rightarrow 2NH_3 \tag{7}$$

$$CO_2 + 2H^+ + 2e^- \rightarrow CO + H_2O \tag{8a}$$

$$CO_2 + 2H^+ + 2e^- \rightarrow HCOOH \tag{8b}$$

$$CO_2 + 6H^+ + 6e^- \rightarrow CH_3OH + H_2O \tag{9}$$

$$CO_2 + 8H^+ + 8e^- \rightarrow CH_4 + 2H_2O \tag{10}$$

$$2CO_2 + 8H^+ + 8e^- \rightarrow CH_3COOH + 2H_2O \tag{11}$$

$$3CO_2 + 18H^+ + 18e^- \rightarrow CH_3CH(OH)CH_3 + \frac{5}{2}H_2O \tag{12}$$

The formation of various other $CO_2RR$ products have been also documented, such as ethanol, ethylene and other hydrocarbons, oxalic acid, acetone and methyl formate.[69] There is currently a large scientific interest on these multi-carbon $CO_2$ reduction reactions as a valuable route to produce high-value chemicals in substitution of current routes from fossil fuels.[70–77] The rich electrocatalytic chemistry of $CO_2$ reduction opens new possibilities to build a new chemistry on it, although understanding how to control the reaction selectivity is still an issue.[78] For conciseness, we will limit our discussion only to few type of $CO_2RR$, which are relevant for SFs used as renewable energy storage system or as hydrogen vector.

The NRR is a reaction of recent large interest, with over 50 reviews on this topic published in the last few years, mainly focused on the electrocatalytic synthesis and related materials, a selection of which is given in ref.[78–97] More limited studies deal with the photochemical NRR, while essentially no studies on PEC-type devices for NRR have been reported. The interest on this reaction derives from the possibility to reduce up to 90% the carbon footprint with respect to the current method of synthesis (Haber-Bosch process), which meets the current need for large quantities of ammonia, a key component towards fertilizers and as hydrogen or energy vector. Large initiatives are current on-going on ammonia as $H_2$ vector, although produced by thermo-catalytic process with $H_2$ by electrolysis as feed.

There is an intrinsic and relevant difference between OWS and $CO_2RR$ and NRR. In fact, eq. 6 (HER) has the great advantages over the others that formally no competitive reactions occur, thus, differently from $CO_2RR$ and NRR, selectivity does not represent an issue. This is especially important in electrocatalysis because the reaction rate could be improved by increasing the applied overpotential, up to the maximum determined by limitations due to ohmic and concentration polarization, mass transport losses, activation losses. When a selectivity issue is present, the overpotential to be applied is instead determined from the compromise between reaction rate and selectivity. Productivity (and optimal $J$) is thus typically lower than in the OWS case.

## Factors limiting the current density

In the previous sections it was shown that solar-to-fuel conversion is a multi-step process, comprising many mechanisms which take place at different timescales. The following sections are dedicated to an analysis of the physical and chemical implications of each process and how they affect the output $J$ of devices for SFs production.

### Photon absorption and generation of free charge carriers

The absorption onset of semiconductors depends on their energy bandgap, $i.e.$, the energy difference between VBM and CBM. The absorption coefficient ($\alpha$) is a measure of how efficiently the incident radiation is absorbed. Consequently, good light harvesters must show panchromatic (to increase the part of absorbed sunlight) and enough thickness to capture all the absorbable light. A high $\alpha$ value has a twofold advantage: reduces the amount of required photo-material (thus, the costs of fabrication) and limits the distance that charge carriers must travel to reach the interfaces of the light harvester. A comparison of $\alpha$ for different photo-materials is shown in Figure 4a, revealing that the absorption onsets usually are not sharp and display sub-band gap absorption because of energetic disorder arising from the presence of defects, trap states and other aspects.[104]

However, high $\alpha$ values do not straightforwardly imply that a photo-material can produce high $J$ values. The ability of a photo-material to generate electron-hole pairs is affected by other processes accounted by the incident-photon-to-current-collected-efficiency ($IPCE$), or external quantum efficiency, defined as:[43,45]

$$\text{IPCE} = \eta_{LHE}\eta_{sep}\eta_{col} \tag{13}$$

Here $\eta_{LHE}$ represents the light harvesting efficiency (which is typically identified with the absorbance), $\eta_{sep}$ is the efficiency of electron-hole pairs separation and $\eta_{col}$ is the probability that electrons and holes reach the interfaces of the light harvester. The product $\eta_{sep}\cdot\eta_{col}$ identifies the flux of collected charge carriers per absorbed photon which is termed the absorbed photon-to-current conversion efficiency (APCE) or internal quantum efficiency.[43,45] The photo-produced $J$ arises from the direct use of solar photons, one-by-one, to generate electron-hole pairs.

The maximum attainable values of photo-produced $J$ (which equals the short circuit current density $J_{sc}$),[99] can be obtained by multiplying the elementary charge $e$ by the number of photons which successfully produce collected charge carriers. In mathematical terms, $J_{sc}$ results from the integral, over all the emitted photon energies, of the solar spectral photon flux $\varphi(E)$ times the $IPCE$:[99]

$$\text{J}_{sc} = e\int_0^\infty \text{IPCE} \cdot \varphi(E)dE \tag{14}$$

In the ideal case where $IPCE = 1$, eq. 14 can be used to calculate the maximum theoretical $J_{sc}$, by using the AM1.5 spectral features in $\varphi(E)$. The resulting $J_{sc}$ is shown in Figure 4b, as a function of the photo-material $E_g$. It is worth noting that $J_{sc}$ increases with decreasing $E_g$ (maximum 73 mA·cm$^{-2}$ for $E_g = 0$ eV), confirming that small band gaps are needed to achieve high $J$. However, charge carriers with energy exceeding that of the band edges ($i.e.$, hot carriers) lose their energy firstly by scattering events with other carriers (thermalization) and lastly by the emission of phonons (cooling).[105] The relaxation of charge carriers to the band edges occurs with time scales ranging between $fs$ to $ns$ (depending on the occurring mechanisms).[106] Actually, there is a large interest in hot carriers properties for many applications such as photodetection,[107] lasing[108] and photocatalysis.[109] However, a thorough discussion of these aspects are beyond the scope of this review. The thermodynamic restrictions due to SFs production pose a sever limit to the maximum reachable $J_{SC}$ (for example, in OWS, a minimum photo-voltage of about 1.6-1.8 eV is required to evolve $H_2$ and $O_2$ with a sufficient rate which correspond to ~25 mA·cm$^{-2}$).

The efficient generation of charge carriers does not depend only on $\alpha$, but on $\eta_{sep}$ too. In fact, when a photon is absorbed, an electron and a hole are generated at the same location and can attract each other through their electrostatic interaction.[110] It is then possible for them to form a neutral bound state ($i.e.$, an exciton), thus reducing the number of free charge carriers contributing to $J$. Excitons can have binding energies ($E_b$) ranging between tens (Wannier-Mott excitons) to thousands (Frenkel excitons) meV.[110] Since thermal energy at room temperature is ~25 meV, Wannier-Mott excitons can be readily dissociated, leading to the generation of free charges. Wannier-Mott excitons can be treated as hydrogenic systems with quantized energy levels given by:[103]

$$\text{E}_b(n) = -\frac{m_r^* e^4}{8h^2\varepsilon_r^2 n^2} \tag{15}$$

where $m_r^*$ is the reduced mass of the electron-hole pair $m_r^* = ((m_e^*)^{-1} + (m_h^*)^{-1})^{-1}$, $h$ is Planck's constant, $n$ is the principal quantum number (for $n=1$, the $E_b$ value is obtained) while $\varepsilon_r$ is the relative dielectric function of the medium surrounding the exciton. The exciton energy levels strongly depend on how the electron-hole pair coulombic interaction is screened by the medium ($i.e.$, $E_b(n)$ is a function of $\varepsilon_r$, as shown in eq. 15). The $\varepsilon_r$ describes the dynamics of the medium and can span over a wide range of frequencies (from $10^4$ to $10^{15}$ Hz), depending on the resonances such as internal molecular vibrations, polar oscillations, internal diffusion of ions and other aspects. This is shown in Figure 4c where the real part of $\varepsilon_r$ is reported for the case of the perovskite $CH_3NH_3PbI_3$.[102]

However, efficient dielectric screening may arise only if the dynamic responses have energy close to that of $E_b$ which, for the case of Wannier-Mott excitons falls in the frequency interval ~ 2-20 THz. This frequency range pertains lattice vibrations arising from phonon modes, so the proper engineering of the vibrational dynamics (*i.e.*, the careful choice of the lattice chemical composition) is a useful tool for tuning $E_b$ and consequently for boosting $J$.

**Transport of charge carriers**

After the absorption of radiation and the formation of free electron-hole pairs, charge carriers must travel through the bulk semiconductor to be collected at the interfaces. The quantity of charges that reaches the contacts gives $J$, so high values can be obtained only with optimal transport properties of the semiconductor. The motion of electrons and holes is hindered by many aspects which can be conveniently grouped in:

1. diffusion/drift factors
2. recombination factors

**Diffusion/drift factors.** The motion of charge carriers in the bulk of semiconductors can be due to a gradient of electron/hole concentrations (called diffusion and described by the diffusion coefficients $D_n$ or $D_p$) and/or by an electric field (called drift and described by the mobilities $\mu_n$ and $\mu_p$). Drift and diffusion are related to each other through the Einstein relation:[111]

$$\mu = \frac{eD}{k_B T} \qquad (16)$$

where $k_B$ is Boltzmann's constant and $T$ is the absolute temperature (eq. 16 holds for both electrons and holes by substituting the corresponding values of $\mu$ and $D$). The factors which limit the transport of charge carriers can be easily identified by considering the definition of $\mu$ on the basis of the Drude model.[112] In this framework, the transport of electrons and holes is hindered by scattering events which randomize the charge carriers flow within the material. The average time between two scattering phenomena is called relaxation time $\tau$ and the mobility can be expressed as:[112]

$$\mu = \frac{e\tau}{m^*} \qquad (17)$$

where $m^*$ is the effective mass of the considered charge carrier (electron or hole). Each material will show different $\mu$ since both $m^*$ and $\tau$ strongly depend on the nature of the investigated material. Indeed, $m^*$ is inversely proportional to the curvature of the CB or VB (depending on the considered charge carrier), while $\tau$ relates to the underlying scattering mechanisms. Thus, $\mu$ can be potentially tuned by optimising $m^*$ (*i.e.*, by varying the electronic energy bands with proper engineering of the chemical composition) and $\tau$. For the latter, scattering may occur because of other charge carriers, imperfections in the crystal (*i.e.*, grain boundaries, impurities, energetic disorder, etc.) and interactions with phonons.[112,113] Scattering from imperfections can be potentially avoided (or at least largely reduced) through proper material fabrication routes. On the contrary, phonon scattering is a fundamental process which can only be partially mitigated. In general, there are two main phonon scattering mechanisms which limit $\mu$: deformation potential and local polarization.[112]

The former affects all materials, since it results from deformations (*i.e.*, variations) of the crystal potential because of lattice vibrations. This phenomenon can arise from both longitudinal and acoustic phonons. The local polarization mechanism is, instead, experienced by polar and partly ionic materials.[112] Such scattering event is due to anti-phase oscillations of oppositely charged ions, which generate a dipole moment and consequently an electric field which affects charge carriers mobilities. In this case, acoustic and longitudinal optic (LO) phonons can contribute to this phenomenon (through the so called piezoelectric acoustic scattering and Fröhlich interaction, respectively).[114] The transverse optic (TO) phonon does not contribute to Fröhlich interaction because it does not produce a macroscopic polarization of the lattice. Since phonon dispersion curves depend on the chemical nature of the crystal, tuning phonon properties is a fundamental strategy to increase $\mu$ (and therefore $J$).

Note that these effects depend also on phonon and carrier confinement due to the dimensionality of the semiconductor.[115] There is thus a dependence on the nanostructure of the materials, which determines the transport of carriers and photocatalytic behaviour. Note also that defects associated to strains created at the interface between grains, for example anatase and rutile phases of $TiO_2$, have a significant effect on phonon lifetime.[116]

**Recombination factors.** Photo-excited electron-hole pairs (with concentration $n$) move through the semiconductor and may recombine (through several mechanisms) reducing the current flow. Inorganic semiconductors such as Si and GaAs show three main recombination pathways: monomolecular, bimolecular and Auger (schematically represented in Figure 4d).[99,117]

The total rate of recombination can be written as:[45,111,117]

$$R_T = R_m + R_b + R_A = An + Bn^2 + Cn^3 \qquad (18)$$

where $A$, $B$ and $C$ are, respectively, the monomolecular, bimolecular and Auger recombination constants. The time evolution of $n$ can be written as:

$$\frac{dn}{dt} = G - R_T = G - An - Bn^2 - Cn^3 \qquad (19)$$

where $G$ is the generation rate of photo-carriers. Extracting the values of $A$, $B$ and $C$ (for example from time resolved spectroscopies)[118] gives fundamental information about the recombination mechanisms occurring in the photo-material. This kind of information is fundamental to boost $J$, as recombination processes result in the reduction of free charge carriers within devices for SFs synthesis.

From eq. 19 another parameter can be introduced, the average recombination lifetime $\tau_{rec}$, *i.e.*, the average time in which photo-generated electrons or holes can move within the semiconductor, before recombining:[98,119]

$$\tau_{rec} = \frac{R_T}{n} = A + Bn + Cn^2 \qquad (20)$$

Finally, it can be shown that the average length travelled by charge carriers before recombining, which is the diffusion length $L_D$, is given by:[106]

$$L_D = \sqrt{D\tau_{rec}} \qquad (21)$$

Thus, $L_D$ gives several information about the semiconductor features, as it depends on both the mobility (through $D$) and the recombination mechanisms. It's worth noting that knowledge of $L_D$ is fundamental for photo-materials since there is a trade-off in the choice of their thickness. High thicknesses increase the probability of photon absorption but thicknesses exceeding $L_D$ will cause materials with potentially high recombination rates (and consequently low $J$).

Table 1 presents the values of several parameters herein introduced, for many benchmark semiconductors used as light harvester in both photovoltaics (PV) and SFs production devices. It is evident that PV-grade materials (i.e., Si, III-V semiconductors, and chalcogenides) exhibit the highest values for both $\mu$ and $L_D$. However, these materials usually suffer from degradation in the electrolyte environment, so they are poorly stable. Quite differently, oxide materials are far more resistant to electrolyte corrosion, but they show poorer charge carrier transport properties, leading to lower $J$ values.[120]

**Table 1** Comparison of the energy band gap ($E_g$), electron ($\mu_e$) and hole ($\mu_h$) mobilities and diffusion length ($L_D$) of semiconductors used as light harvesters in PV and SFs devices.

| Material | $E_g$ eV | $\mu_e$ cm$^2$ V$^{-1}$ s$^{-1}$ | $\mu_h$ cm$^2$ V$^{-1}$ s$^{-1}$ | $L_D$ $\mu$m | Ref |
|---|---|---|---|---|---|
| c-Si | 1.12 | 1450 | 500 | 100 | 121,122 |
| GaAs | 1.42 | 8500 | 400 | 100-900 | 121,123 |
| CuInGaSe$_2$ | 1-1.7 | 100 | 2.5 | 0.3-0.52 | 120,124 |
| Sb$_2$Se$_3$ | 1.1-1.2 | ~10 | ~10 | 0.28 | 100,125 |
| TiO$_2$ | 3.02 | 0.01 | 100 | 0.01-0.1 | 120,126,127 |
| WO$_3$ | 2.8 | 10 | 20-104 | 0.15-0.5 | 120,128,129 |
| Fe$_2$O$_3$ | 1.9-2.2 | 0.5 | 10$^{-7}$-10$^{-4}$ | 2-4 10$^{-3}$ | 120,129,130 |
| Cu$_2$O | 2.14 | 6 | 256 | 0.025 | 120,131,132 |
| BiVO$_4$ | 2.4 | 0.044/12 | 2 | 0.07 | 120,133,134 |
| Ta$_3$N$_5$ | 2.01/2.12 | 1.3-4.4 | 0.08 | 18 | 120,135,136 |
| TaON | 2.08/2.4 | 17-21 | 0.01 | 8 | 120,135,137 |
| CH$_3$NH$_3$PbI$_3$ | 1.6 | 1.4 | 0.9 | 0.3-1 | 98,138 |

The discussion presented so far describes the processes involved for the generation and efficient separation of charge-carriers inside the light harvester. It must be remembered that when electrons and holes reach the absorber interfaces, they must migrate through them and eventually through other layers (used as protective barriers or to facilitate charge transport) before reaching the active sites (for the reaction of interest) or the back contact. To ensure the production of high $J$, it is essential to minimize both resistance to transport and possible charge recombination, both highly depending on the interface present. All components of a device must be carefully optimized from this perspective, although often this aspect is not accounted in simple device configurations. Thus, while the light harvester is essential to both generate and transport efficiently electrons and holes, all the other device components (such as transporting, passivating or buffer layers and contacts) must guarantee a low recombination probability of charge carriers, to prevent photo-current loss.

Nanostructuring the semiconductor is relevant from this perspective.[139] For example, a 1D-type ordered nanostructure (as in TiO$_2$ ordered arrays of nanotubes)[140] allows to realize a vectorial transport with holes-electrons moving on the internal or external sides of the nanotubes, reducing thus the possibility of recombination.[141] However, this positive effect is often masked from the presence of shallow trap states that instead favour recombination and reduces charge-carrier mobility.[142] A proper mastering of all these factors is thus crucial to enhance the device performances.[143]

### Redox (catalytic) reactions

Once electron-hole pairs are formed, separated, and transported to the interfaces of the absorber, they must reach the back contact or the active sites (at the interface with the electrolyte) to trigger oxidation or reduction reactions. However, these reactions can take place only if the electronic band edges straddle the redox levels, i.e., the CBM must lay above the reduction potential while the VBM must lie below the oxidation potential.

This condition would pose a further limit to the choice of light harvesting materials, which now must realise these conditions: (i) maximise the part of absorbed light, (ii) produce enough photovoltage and (iii) show proper band edges energetics with respect to the desired reactions. Since finding a material that can meet all these requirements is particularly challenging, usually the processes of charge carrier generation and electrocatalysis are decoupled by using an optimised light harvester (which photo-produces carriers) and electrocatalysts (to made the considered chemical reactions).

The optimization of the electrocatalysts properties is pivotal to achieve high $J$, as many processes may hinder charge carrier motion. First, electrons or holes must be driven efficiently from the light absorber/electrocatalyst interface to the electrocatalyst/electrolyte interface, so the transport must be optimal to avoid loss due to recombination effects at both absorber/electrocatalyst interface and within the electrocatalyst itself.

Moreover, when charge carriers reach the electrocatalyst/electrolyte interface they can finally trigger the chemical reactions, but if the carrier transfer (from the electrocatalyst to the reactant/electrolyte) is too slow, the chances of recombination events increase, leading again to a reduction of $J$. Such charge transfer can be quantitatively assessed through the exchange current density, which is another fundamental parameter that must be considered when systems for SFs production are designed. The exchange current density can be also related to the chemisorption energy of the products on the cocatalyst surface, as observed by Trasatti for the case of metallic cocatalysts.[144] The resulting trend resembles a triangle, thus the name given to such shape is "volcano plot". Note that this trend is based on several assumptions (for example that the exchange current depends on the strength of the M–H bond), and thus that various experimental results deviate from this trend.[145] In oxide electrodes, where large deviations from the outer sphere electron transfer mechanism are present (larger than in metallic electrodes), deviations from the predicted volcano-type trend are large.[145] Moreover, in complex reactions, such as CO$_2$RR and NRR, volcano-type approaches fail in giving correct predictions. Cocatalysts are often used to improve the redox reaction rates, but their design is largely phenomenological, and a better theoretical fundamental approach for their rational design would be necessary.[146–148]

In practical terms, a methodology largely used for the investigation of these aspects is the Tafel plot,[149] which is an experimental measure of the current density (usually in a logarithmic form) and the applied overpotential. The onset of the curve defines the exchange current density. The Tafel equation approximates the Butler-Volmer equation, when the concentrations at the electrode are equal to the concentrations in the bulk electrolyte, allowing the current to be expressed as a function of potential only.

Finally, electrocatalysts should exhibit long-time operational stability, *i.e.*, they should maintain their performances unchanged for long-time use at high $J$. In fact, instability arises from many processes such as oxidative decomposition, metal leaching and peeling of the electrocatalyst from the photoelectrode. Currently, electrocatalysts exhibit a stability of tens of hours (while very few survive for more than hundreds of hours) at $J \sim 10$ mA·cm$^{-2}$, while for practical applications current densities of at least one order of magnitude higher are requested (commercial electrolysers work at $J \sim 0.5$-2 A·cm$^{-2}$).[39,40] A very encouraging and interesting result about long-living electrocatalysts recently came from layered double hydroxides, which show an excellent and stable catalytic activity for over 6000 hours at $J = 1$ A cm$^{-2}$,[150] which represents (to the best of our knowledge) the current record for highly stable and efficient electrocatalysts.

**Mass transfer**

Other aspects that need to be considered in electrochemical processes are the limitations related to mass transfer, related to the ion mobility in the electrolyte (which consists of ionic active species dissolved in a solvent). For OWS, the active species coincide with the solvent, *i.e.*, water. However, the resistivity of pure water is high (distilled water $\sim 10^5$ Ω, purified water $\sim 10^7$ Ω),[37] so supporting ions are usually introduced to avoid Ohmic losses throughout the electrolyte environment. The potential drop due to the electrolyte resistivity is given by:

$$V_{loss} = i \cdot R_E \qquad (22)$$

where $i$ is the current flowing between the working and counter electrodes, while $R_E$ is the electrolyte resistance, defined as:

$$R_E = \frac{K_{cell}}{\sigma} \qquad (23)$$

where $\sigma$ is the electrolyte conductivity and $K_{cell}$ is the cell constant, which depends on the reactor geometry.

Such resistivity influences the transport of the ions through the solvent. These transfer processes are intrinsically limited by the electrolyte properties and can be described by considering three mechanisms: (i) diffusion, due to chemical potential gradients, (ii) migration, induced by electric fields and (iii) convection, resulting from the electrolyte motion. Mass transport is usually described theoretically through the Nernst-Planck equation, which is given, for the $k$ species along the x-direction (one-dimensional case) by:[48,151]

$$J_k(x) = -D_k \frac{\partial a_k(x)}{\partial x} - \frac{z_k F}{RT} D_k a_k \frac{\partial \phi_k(x)}{\partial x} - a_k v(x) \qquad (24)$$

where $J$ is the flux, $D$ is the diffusion coefficient, $a$ is the activity, $z$ is the charge number, $F$ is the Faraday constant, $R$ is the universal gas constant, $\phi(x)$ is the electric potential and $v(x)$ is the velocity vector of the fluid. The right-hand side comprises three terms which refer, respectively, to diffusion, migration, and convection phenomena. It is worth mentioning that the diffusion coefficient can be expressed, through the Stokes-Einstein relation as:[151]

$$D = \frac{k_B T}{3 \pi d v} \qquad (25)$$

where $d$ is the diameter of the hydrated ion (Stokes diameter) and $v$ is the viscosity of the solution. It is then clear that several physico-chemical properties must be considered to optimise mass transport in the electrolyte (resistivity, activity and dimension of the ions, viscosity of the solution and other aspects) to optimize the performances in terms of $J$, scalability, and industrialization of the devices.[152–154] Fluidodynamic modelling of the electrolyte in the PEC cell is complex due to the presence of an electrical field, a (typically) multicomponent porous electrode, the formation of gas bubbles and their adhesion to the electrode, the possible crossover over the membrane and other effects, but is a crucial element to optimize PEC and PV/EC cell engineering and scale-up.[154]

Moreover, CO$_2$RR and NRR suffer from limitations related to the low solubility of CO$_2$ ($\sim 3 \cdot 10^{-2}$ M)[155] and N$_2$ ($\sim 7 \cdot 10^{-4}$ M) in H$_2$O, that intrinsically limits the maximum achievable $J$ in the range of tens mA·cm$^{-2}$. For this reason, other configurations have been investigated that favour high-$J$ working conditions, such as pressurized electrolyte systems and gas-diffusion-layer (GDL) architecture.[156] In particular, in GDL-based devices the electrodes are in contact directly with the membrane acting as separation element between the two cells, but also as element to provide the ionic closure of the circuit.[28,139,157,158] The photo- and electro-catalysts are present in the GDL electrodes put in contact with the two sides of the membrane.[28,157] Among the advantages of this configuration (called electrolyte-less or gas-phase design) are the (i) not-energy-intensive recover of the reaction products, (ii) easier scale-up and lower costs, (iii) more compact design, and (iv) elimination of the problems of reactant solubility in the electrolyte and diffusional limitations through the double electrical layer. Due to the latter aspect, the electrodes operate in the presence of different surface coverages of the reactants, giving rise to different products in CO$_2$ electrocatalytic conversion, for example.[158]

This concept is presented schematically in Figure 5, where the comparison between the architectures with (Figure 5a) and without (Figure 5b) electrolyte is presented (in this latter case, the closure of the electric circuit is provided from the membrane itself, which thus acts also as a solid electrolyte).[28]

The absence of a liquid electrolyte makes scale-up easier, and cell manufacture cheaper. Corrosion issues and material stability are largely overcome. However, mass transfer limitations in the two PEC cell design are significantly different, due either, in one case, to the absence of a liquid electrolyte, and in general to the different transport mechanisms for the two configurations.

In the compact design, oxygen evolution occurs by water oxidation (eq. 26a), while hydroxyl oxidation (eq. 26b) occurs at the anode for the conventional cell design of Figure 5a. Note that in the latter case, protons migrate through the membrane separating the two half-cells (Figure 5a), while in the compact design protons migrate firstly through the porous oxide semiconductor, then through the membrane between the anodic and cathodic gas-diffusion-layer type electrodes.

$$2H_2O + 4h^+ \rightarrow O_2 + 4H^+ \qquad (26a)$$

$$2OH^- + 4h^+ \rightarrow O_2 + 4H^+ \qquad (26b)$$

The compact design PEC cell requires to have an oxide semiconductor nanostructure allowing surface diffusion of protons (like based on ordered arrays of vertically aligned oxide nanotubes).[159,160]

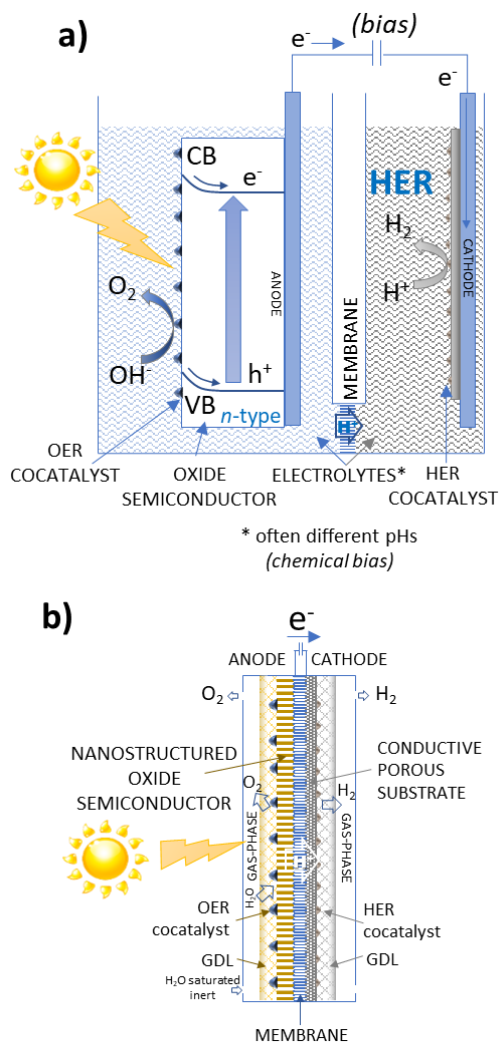

**Figure 5.** a) Scheme of a conventional PEC cell operating with the two electrodes immersed in a liquid electrolyte. b) Scheme of a compact (electrolyte-less) design for PEC operations.

## The impact of the device architecture

Herein, the discussion will focus on PEC and PV/EC design, which represent currently the most promising technologies for SFs generation. As remarked before, the case of PP architecture is not addressed here, being not possible to measure a $J$. In addition, this architecture still shows low efficiencies. For example, Domen and collaborators reported recently a Z-scheme system (La- and Rh-codoped SrTiO$_3$ and Mo-doped BiVO$_4$ powders embedded into a gold layer) as a superior system for OWS, although still having efficiency of 1.1%.[161] We want to emphasise that our aim is to give a viewpoint from which new considerations and studies can be carried out. We will present only some relevant results about these topics since making a detailed state-of-the-art review is out of the scope.

The production of SFs requires the use of energy which can be harvested by sunlight (unassisted SFs production) or be partially provided by an external bias (assisted SFs production). From a practical and cost perspective, only the first case is relevant and could be used as first generation artificial-leaf devices. Note that a device having one light harvester (without sun concentrators) produces at the best $J_{SC}$ <70 mA·cm$^{-2}$ (*i.e.*, the maximum photo-generated $J$; which depend on the material's properties). This holds true for both photoelectrodes and solar cells, so PEC and PV/EC designs. The advantage of PEC devices is that they are well-compact integrated systems (*i.e.*, they can both harvest light and make chemical conversion), thus potentially having reduced cost of SFs production.

In addition, as commented later, they are better suited for use in conjunction with sun concentrators than PV/EC devices. On the other hand, there are various other aspects to consider which are in favour of PV/EC design. For example, the substantial projected cost of running electrolyte over large areas and providing non-fouling transparent flow channels for these devices which operate at low $J$. Also joined multi-junction solar cells are currently the most efficient PV devices. As discussed later, PV/EC cell design is that currently giving the best performances, but a detailed techno-economic comparative analysis of the different cell configurations, and their pro/cons is still missing in literature.

In terms of industrial use, many issues must be assessed, such as: (i) the development of efficient absorbers at low cost and from large-scale fabrication methods,[162] (ii) increasing stability especially towards photocorrosion,[152] (iii) improving the performances. These criteria translate into the great challenge of finding optimized materials and device architectures.[163] With the aim to increase $J$, the $E_g$ of the photo-absorber must be small thus decreasing the generated photovoltage, but which may result too small to trigger the redox reactions. In this case, an external bias is needed (assisted SFs production).

Unassisted SFs production can be realised through PC/PA architecture, when both are photoactive. By connecting the two photoelectrodes in series, the resulting voltage will be roughly the sum of the photovoltages produced by the PC and PA components while $J$ will be the same for both (*i.e.*, it will equal the smallest produced $J$). There are several possible reactor designs for PC/PA systems, depending on both the illumination setup and the photoelectrodes assembly.[164] The simplest model is the one introduced in the previous sections, with PC and PA facing each other and connected by wires (Figure 3d), so the front photoelectrode needs to be transparent. Another interesting architecture is the monolithic back-to-back stack of both PC and PA, separated by a charge collector, resembling a

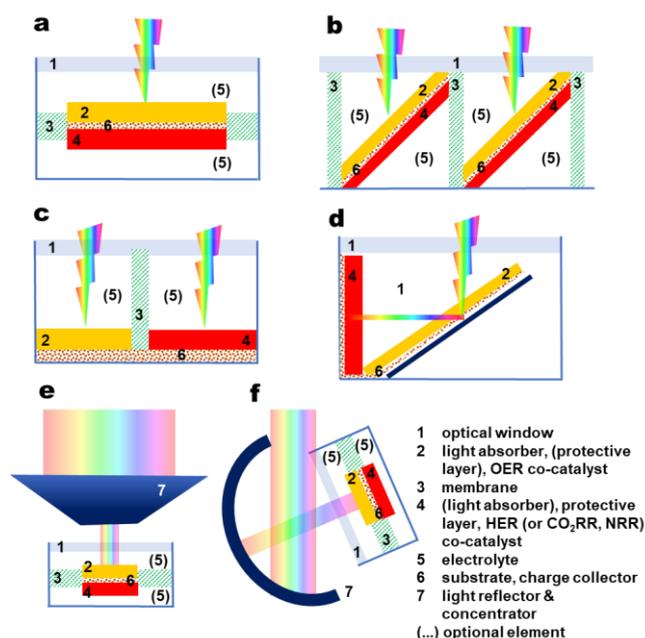

**Figure 6.** Architectures for PEC systems. Monolithic a) planar and b) tilted PC/PA configuration. Side by side configuration of c) planar and d) tilted PC and PA electrodes. e-f) Monolithic PC/PA devices coupled with sunlight concentration systems. Re-elaborated from ref.[9]

Legend:
1 optical window
2 light absorber, (protective layer), OER co-catalyst
3 membrane
4 (light absorber), protective layer, HER (or $CO_2RR$, NRR) co-catalyst
5 electrolyte
6 substrate, charge collector
7 light reflector & concentrator
(...) optional element

solar panel dipped in the electrolyte (with the eventual variation of electrolyte-less configuration discussed previously) as shown schematically in Figure 6a. For this wireless configuration (named "artificial leaf" by Nocera),[32] there is no ohmic loss associated to wires, but good transparency is needed for photoelectrodes and charge collector, limiting considerably the material selection. Such wireless configuration shows lower performances than the wired counterparts, because of poor charge transfer and ohmic contacts between the different layers.[9,164] A tilted configuration can optimise the spatial arrangement (Figure 6b), moreover the PC and PA can also be put side by side (Figure 6c), avoiding the need for transparency but doubling the space requirements. A tilted configuration is also possible (with PC facing PA at 45°),[165] but it makes difficult product separation (Figure 6d).[166] Thus, whatever PC/PA series architecture is considered, the highest $J$ is always limited to ~70 mA·cm$^{-2}$. Only the parallel connection of two photoelectrodes results in a $J$ which equals the sum of the $J$ produced by the single photoelectrodes.[167] However, the output voltage would be almost equal to the mean value of the voltage produced by the two photoelectrodes, so an external bias would be needed to reach the minimum potential to drive the reaction process.

Note finally that if materials are robust enough, PEC cells could be integrated with light concentration elements (Figure 6e-f). Here, higher $J$ can be obtained (since in eq. 14 the intensity of $\varphi(E)$ won't be that of one SUN illumination - ~100 W·cm$^{-2}$ - but higher), according to the level of sun concentration.[168] For PV cell, the effectiveness in increasing the $J$ by solar concentrations is limited from the negative effect of the increase of the temperature on the PV cell performances. However, using photoelectrodes in PEC cell makes this limitation likely less relevant (the dependence from the temperature of the performances of oxide semiconductor photoanodes is different from that of PV cells), although thermal robust components for the PEC cell are necessary. This is a largely unexplored area, but crucial to enhance PEC $J$ and performances.

Materials and cells able to operate in a medium temperature range (50-200°C), which is not compatible to most of existing technologies, should be developed. However, the drastic possible reduction in costs (which largely depend on $J$, being a main parameter determining cell productivity) is a main driver from an industrial perspective.

A related challenge is that the devices should operate efficiently also at variable operational temperatures, because of fluctuations of the SUN intensity due to seasonal variations as well as geographic location. It is thus an area which can open new application perspective, but which is still largely unrecognized.

As concerns the large scale implementation of PEC systems, many practical issues must be properly assessed such as performance of scaled-up electrodes (which may reach ~80% loss of current with respect to lab-scale devices and increased ohmic resistances,[9] although the effective loss of performances could be lower than 10%), the need of avoiding critical raw materials for the realization of the PEC system and especially cost-effective solutions (including stability in the analysis) and safety concerns for operations. While commercialization of the systems for OWS is expected within 5-10 years, those for $CO_2RR$ and NRR will take longer time for the higher challenges in their development.

The other main architecture for SFs production is the PV/EC assembly, which presents several practical advantages with respect to other approaches. First, the carrier generation/ transport and catalytic conversion processes are decoupled, so they can be optimised by the proper choice of the PV and EC components. Many PV technologies are already commercially available, each one being characterised by different performances (in terms of PV parameters such as $J_{SC}$, open circuit voltage, power conversion efficiency and other aspect), stability and cost.[99,167] Furthermore, in such architecture the PV component lies outside the electrolyte environment, avoiding detrimental effects due to electrochemical corrosion. Finally, it has the potential to deliver high $J$ by connecting in parallel more solar cells. The resulting current will be the sum of all the currents produced by the single devices. However, the working point of the PV/EC assembly equals the intersection of the $J$-$V$ curves of the PV and EC components, as shown schematically in Figure 7 (thus the power input to the EC equals the power output of the PV).[169]

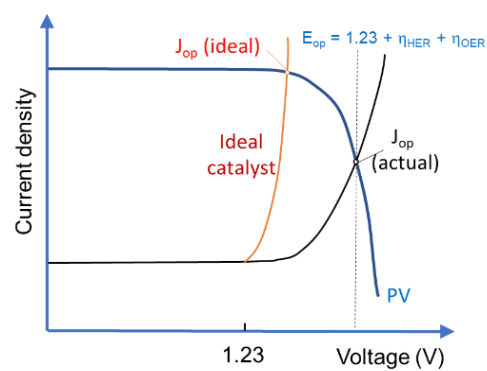

**Figure 7.** Current density-potential behaviour for a photovoltaics and a water electrolyser. Elaborated from ref.[170] Copyright Elsevier @2017

For an ideal system, the intersection would occur at the maximum power point of the PV unit ($J_{op}$ ideal, reported in Figure 7), but the overpotential associated to the considered reaction can shift such point to lower $J$ ($J_{op}$ actual, shown in Figure 7). In the PV/EC design $J_{SC}$ indicates the photocurrent produced by the PV unit, while $J_{op}$ refers to the practical current exploited during the electrochemical process. Thus, a correct definition, not always present, is fundamental.

Concentrator-based PV/EC systems (using low-cost Fresnel lens) could deliver both high-$J$ and $\eta_{STF}$. Tembhurne et al.[171] designed a device (for OWS) capable of reaching $J$ up to 0.88 A·cm$^{-2}$ at $\eta_{STH}$ = 17.12%. As early commented, PV performances are sensitive to increase of the temperature and thus the reported performances were obtained by using a thermal management approach that synergistically cools down the PV unit, while warming up the electrocatalytic components of the EC. This is a promising strategy towards practical and large-scale SFs, although likely a better engineering of the thermal management component (and its better integration in the cell) will be necessary.

It is worth mentioning that the PV/EC architecture is, to the best of our knowledge, the only one already evaluated on a large scale, at least for H$_2$ production.[9,172–174] Maeda and Domen estimated that with 25·10$^4$ km$^2$ of solar power plant (i.e., 1% of Earth's desert land) and $\eta_{STH}$ = 10%, 570 tons of H$_2$ can be produced per day which would represent one third of the 2050 energy demand.[175]

## Status on current density performances

In the following sections, a concise analysis of the discussed architectures to produce SFs in OWS, CO$_2$RR and NRR cases is presented, by using $J$ as a FoM to analyse the data. For OWS, as commented before, the PV/EC approach is currently preferable to reach high-$J$, especially when using low-cost PV cells coupled to sun-concentrators, when thermal heating issue is solved. For CO$_2$RR, and similarly for NRR, the preference between PV/EC and PEC architecture is still an open question.

### The case of OWS

The reduction of H$_2$O to produce H$_2$ is the most studied SFs case. Many literature contributions have discussed the current state of art for H$_2$ production and its potential use.[3,9,21,152,187,188] For this reason, herein we limit the discussion to few representative results which would also serve as a basis for the following considerations on CO$_2$RR and NRR.

A good way to summarize the status of the different approaches used in OWS is presented in Figure 8 elaborated from the original figure presented by Kim et al.[9] They summarized the results of $\eta_{STH}$ in H$_2$ production by OWS as a function of the arbitrary index of system complexity, allowing to cluster together results as a function of the system architecture (PP, PEC, PV/EC). The indicative region (in green) representing that of potential interest from a practical perspective is also outlined. Using this approach, it is possible to have a quick view of the state-of-the-art and of the performances as a function of the device architecture.

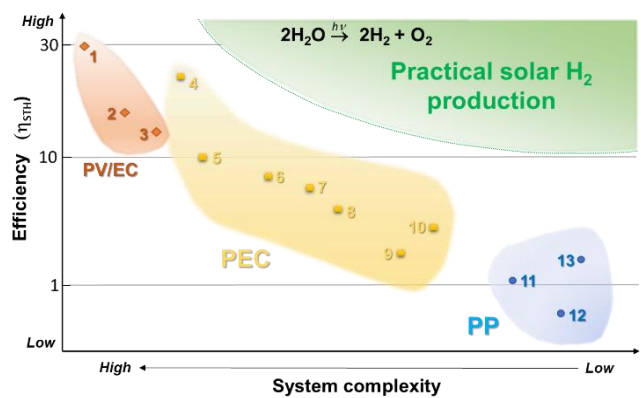

**Figure 8.** Summary of representative results about H$_2$ production through PP, PEC and PV/EC approaches. **(1)** Light concentrator – GaInP/GaAs/GaInNAsSb – PEM, $\eta_{STH}$ = 30% (ref.[60]). **(2)** Si – PEM, $\eta_{STH}$ = 14.2% (ref.[176]). **(3)** Perovskite solar cell – NiFe, $\eta_{STH}$ = 12.3% (ref.[177]). **(4)** GaInP/GaInAs/GaAs, $\eta_{STH}$ = 19% (ref.[178]). **(5)** InGaP/GaAs/Ge, $\eta_{STH}$ = 10% (ref.[179]). **(6)** BiVO$_4$/WO$_3$ – InGaP/GaAs, $\eta_{STH}$ = 8.1% (ref.[165]). **(7)** BiVO$_4$–Fe$_2$O$_3$ – 2pSi, $\eta_{STH}$ = 7.7% (ref.[180]). **(8)** BiVO$_4$/perovskite, $\eta_{STH}$ = 6.2% (ref.[181]). **(9)** BiVO$_4$ – p-Si, $\eta_{STH}$ = 2.0% (ref.[182]). **(10)** BiVO$_4$ – Cu$_2$O, $\eta_{STH}$ = 3.0% (ref.[183]). **(11)** Mo:BiVO$_4$ – Au – Rh,La:SrTiO$_3$, $\eta_{STH}$ = 1.1% (ref.[184]). **(12)** Al:SrTiO$_3$, $\eta_{STH}$ = 0.4% (ref.[185]). **(13)** C$_3$N$_4$, $\eta_{STH}$ = 2% (ref.[186]). Elaborated from ref.[9] Copyright Royal Society of Chemistry @2019

The highest performances come from PV/EC systems, where $\eta_{STH}$ ~ 30% has been demonstrated by using a commercially available InGaP/GaAs/GaInNAs(Sb) triple-junction solar cell, at 42 SUNs (delivering a $J_{SC}$ = 584 mA·cm$^{-2}$), connected in series to two polymer electrolyte membrane (PEM) electrolyzers.[60] Although such performances represent a remarkable result, it should be noted that the authors report a $J_{SC}$ of ~14 mA·cm$^{-2}$ at 1 SUN, which is lower with respect to the theoretical limit of the photocurrent for this kind of PV unit. In fact, in multi-junction solar cells (and photoelectrodes as well) the resulting photocurrent is limited by the highest $E_g$ material (because the same current flows in series-connected light harvesters). For the case at hand, the InGaP absorber layer has the highest $E_g$ (1.895 eV) thus $J_{SC}$ is limited to ~17.5 mA·cm$^{-2}$, underlying that this parameter could be further boosted leading also to higher $\eta_{STH}$. Interestingly, the $J_{op}$ of the assembly remains high throughout the experimentation (48 h) going from ~570 to ~506 mA·cm$^{-2}$.

Another recent work highlighted the high potential of the PV/EC architecture in OWS by exploiting a InGaP/InGaAs/Ge solar cell and an electrolyte-less EC based on a IrRuO$_x$-Pt catalyst.[171] The authors reported a $J_{SC}$ of 5.49 A·cm$^{-2}$ (at 474 SUNs) and an impressive $J_{op}$ of 0.88 A·cm$^{-2}$ and a $\eta_{STH}$ = 17.12%. The about 500 times solar concentration requires a proper thermal management of the system as commented before. Indeed, the authors report $J_{SC}$ = 13.9 mA·cm$^{-2}$ which is lower with respect to the expected limit of this kind of multi-junction solar cell. As emerges from the discussion of these two results, it is necessary to integrate results as those presented in Figure

| Design | Components | $J_{SC}$ / $J_{op}$ (mA·cm⁻²) | $\eta_{STF}$ (%) | Ref |
|---|---|---|---|---|
| PV/EC | GaInP/GaAs/GaInNAs(Sb)//PEM (42 SUNs) | $J_{SC}$ = 584 $J_{op}$ ≈ 538 | 30 | 60 |
| PV/EC | InGaP/InGaAs/Ge//IrRuO$_x$-Pt (474 SUNs) | $J_{SC}$ = 5490 $J_{op}$ ≈ 880 | 17.12 | 171 |
| PEC monolithic | Rh//GaInP/GaInAs//RuO$_x$ | $J_{SC}$ = 15.7 | 19 | 178 |
| PEC | Ta$_3$N$_5$/Ni(OH)$_x$ | $J_{SC}$ = 12.1 | 14.88* | 202 |
| PEC | TiO$_2$/CoNi | $J_{SC}$ = 4.4 | 5.41* | 203 |
| PEC | BiVO$_4$/β-FeOOH | $J_{SC}$ = 4.3 | 5.28* | 204 |

8 with proper considerations on the *J*, to derive reliable conclusions on the more promising materials and approaches.

The indication emerging from this combined analysis is that for OWS case, the PV/EC architecture allows to obtain high-*J* and $\eta_{STH}$, especially when coupled to a concentrator system and a proper heat management system, but which increase the costs. Furthermore, performance losses after long-lasting use must be accurately assessed (such as the detachment of catalyst particles and membrane degradation).[60] The introduction of low-cost solar concentration elements coupled to an efficient thermal management of the PV/cell, together with the use of thermal robust materials for the cell itself, will be likely the crucial factor for commercialization, although extremely few papers (if any) of the very abundant literature on the topic address these three key aspects.

Through the years, the PEC architecture was the most investigated, since it would provide (i) 2% < $\eta_{STH}$ < 19%, (ii) easier and cheaper separation of the reaction products and (iii) potential lower costs for the fabrication of the photoelectrodes with respect to the PV/EC assembly. However, to the best of our knowledge, no large-scale demonstration of PEC powered systems has been reported yet.

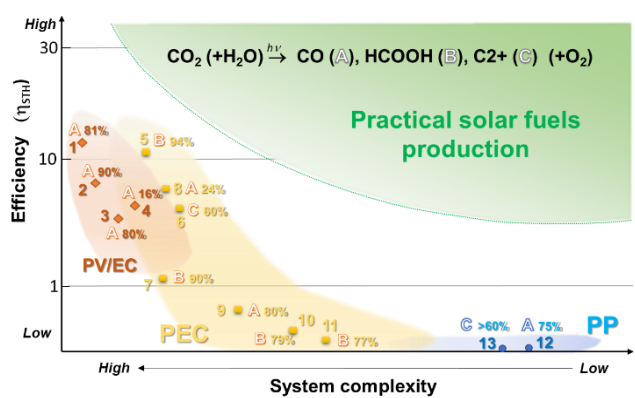

**Figure 9.** Summary of representative results about CO$_2$RR through PP, PEC and PV/EC approaches. The reported values in percentage indicate the Faradaic Efficiencies for the investigated reaction: CO (A), HCOOH (B) and C2+ (C). (1) GaInP/GaInAs/Ge/SnO$_2$-CuO//SnO$_2$-CuO, $\eta_{STF}$ = 13.4% (ref.[189]). (2) Perovskite solar cell/Anodized Au//IrO$_2$, $\eta_{STF}$ = 6.5% (ref.[190]). (3) Si/Mn complex//FeO$_x$-NiO$_x$, $\eta_{STF}$ = 3.4% (ref.[191]). (4) Si/Cu-Zn//Ni, $\eta_{STF}$ = 4.3% (ref.[192]). (5) Pd/C/Ti//Ni/TiO$_2$/InGaP/GaAs, $\eta_{STF}$ = >10% (ref.[193]). (6) Cu/Ag/TiO$_2$//IrO$_2$/perovskite, $\eta_{STF}$ = 3.5% (ref.[194]) (7) CuFeO$_2$/CuO// Pt, $\eta_{STF}$ = >1% (ref.[195]). (8) WSe$_2$/PVaSi3jn//CoOx/PVaSi3jn, $\eta_{STF}$ = 4.6% (ref.[196]). (9) ZnO-ZnTe@CdTeAu// Co-HCO3/Ni$_{foam}$, $\eta_{STF}$ = 0.4% (ref.[197]). (10) Ru complex/ TiFeCrO$_x$//TiO$_2$, $\eta_{STF}$ = 0.15% (ref.[198]) (11) Bi/TiN//FeO$_x$/BiVO$_4$/ perovskite, $\eta_{STF}$ = 0.1% (ref.[199]). (12) Co$_3$O$_4$ platelets- [Ru(bpy)$_3$]Cl$_2$, $\eta_{STF}$ = <0.06% (ref.[200]). (13) Rh-TiO$_2$, $\eta_{STF}$ = <0.04% (ref.[201]).

The best performances have been obtained for tandem PEC systems, mainly based on III-V semiconductors light harvesters. For example, Cheng et al.[178] achieved $\eta_{STH}$ = 19%, at $J_{SC}$ = 15.7 mA·cm⁻², using a monolithic device made up by a GaInP/GaInAs two-junction absorbing layer using Rh and RuO$_x$ as HER and OER catalysts, respectively. Interestingly, the authors performed some calculations to assess the origin of losses in their device. According to their results, kinetic, diffusion and ohmic contributions to *J* and $\eta_{STH}$ do not represent the dominant losses mechanisms. In fact, by measuring the IPCE of their device, the performances were found to be limited by $J_{SC}$, since it should potentially reach ~17.5 mA·cm⁻². Thus, this work emphasises that a careful analysis of the origin of losses in *J* in the device can guide the choice and optimization to maximize the production of SFs. Often these aspects are more critical than the typical aspects related to catalyst design discussed in most of the reviews on the topic. However, there are only few literature results that present such kind of analysis, making comparisons with other studies rather difficult.

For OWS using PEC architecture, several materials have been intensively studied, for example binary oxides, oxynitrides and photovoltaic-grade materials among others. They were discussed in various reviews,[9,21,50,120] while an analysis of their performances is beyond the scope of this review. As mentioned earlier, for all these systems the maximum $J_{SC}$ is limited by the $E_g$ of the material used as light harvester, typically not exceeding ~20 mA·cm⁻² (Table 2). Since for industrial applications a $J_{SC}$ > 300-400 mA·cm⁻² should be indicatively targeted, the PV/EC approach currently represent the only choice to meet this goal, in the absence of using sun-concentration elements. Even for PV/EC approach, this target could be reached only by using sun-concentrations elements as shown in Table 2. There are intrinsic limits both in PEC and PV/EC approaches to reach these targets without a sun-concentration component and the associated system of thermal management, although an optimal trade-off between costs and performances is necessary, as well as considerations on the stability of the systems. On the other hand, it is evident that without an analysis of *J*, these critical aspects in the discussion about solar devices are not emerging.

**Preliminary considerations on CO$_2$-RR and NRR**

The realization of efficient CO$_2$RR and NRR suffer from additional relevant limitations with respect to OWS. In particular, differently from OWS where a single product is formed (H$_2$), a range of chemicals can be synthesised in both CO$_2$RR and NRR. Furthermore, for both CO$_2$RR and NRR the H$_2$ side formation influences the Faradaic selectivity.

For CO$_2$RR, it should be distinguished between carbon Faradaic selectivity (which refers to the selectivity with respect to only the carbon-based products formed) and total Faradaic selectivity, which includes H$_2$. Both are relevant, but carbon Faradaic selectivity is often more important for industrial exploitability. In NRR, other products can be formed in addition to NH$_3$, such as NH$_2$NH$_2$, but their contributions are usually negligible. Thus, total Faradaic selectivity is the typically only considered.

For CO$_2$RR, the resulting products have a different value (in terms of performances as SFs) and a mixture of chemicals can be obtained. This is an aspect often ignored, although from the application perspective of these devices the downstream separation and purification processes could be the determining cost element for implementing the SFs production. Recovery of the products during continuous operations and avoiding electrolyte degradation (also related to by-products accumulation) are other crucial elements scarcely considered in literature.

Finally, the low solubility of CO$_2$ in H$_2$O-based electrolytes intrinsically limits the maximum $J$ to ~35 mA·cm$^{-2}$, about one order of magnitude lower of an indicative value of > 200-300 mA·cm$^{-2}$ to have productivities in the range of industrial relevance.[156]

A starting point to compare with OWS case is to have a graph analogous to that presented in Figure 8, because this kind of comparison is absent in literature. Figure 9 summarises these results for CO$_2$RR case, with an indication of also some of the possible products of reaction. Note that the limits of range of efficiency versus system complexity are larger for CO$_2$RR with respect to H$_2$ production, because the products have in average a higher added value and usability. In addition, they contribute directly to the reuse of CO$_2$ from emissions (in perspective from the air). However, more precise estimations should be made case by case, by considering the specific products formed.

Figure 9 could be used to have an indication about the state of CO$_2$RR with respect to OWS (H$_2$ production). Also note that despite the abundant literature on CO$_2$RR, only few results provide data in terms of reliable indication of $\eta_{STF}$ in the absence of external bias and added sacrificial agents. Thus, Figure 9 can stimulate the readers to provide in future publications data which could be directly put in relation to this graph.

However, a similar figure cannot be made for NRR for the absence of enough reliable data. All results of photocatalytic N$_2$ reduction do not provide indications on $\eta_{STF}$. Those for which an estimation can be made (from available data of external quantum yield) indicate $\eta_{STF}$ significantly lower than 0.1%. Some papers report a PEC or PV/EC approach in NRR, but with an applied external bias. Without this additional voltage (ranging typically between 0.3-0.6V), the formation of NH$_3$ is quite negligible, indicating extremely low $\eta_{STF}$ values. Quantitative indications of $J$ for NH$_3$ production are extremely limited, even for the case of a pure electrochemical approach (thus with external bias) for which a maximum around 0.5 mA·cm$^{-2}$ was demonstrated (with Faradaic efficiency to ammonia below 30%).[88,205]

For unassisted PEC or PV/EC approaches without sacrificial agents, $J$ values are lower, nearly negligible in the few results reported up to now (see the following section for more details). Thus, NRR stage of development is still largely below that for CO$_2$RR and OWS, preventing an analysis as that attempted in Figures 8 - 9.

In addition, in CO$_2$RR and NRR the operational $J$ is determined from the need to limit the side reaction of HER and thus productivity is limited. For CO$_2$RR and NRR, the reaction mechanism is more complex than that for H$_2$ production because more electrons/protons are involved, and reaction intermediates and possible pathways are competitive to the SF production. All these steps should be optimized and will depend on the applied potential. PV/EC approach, which is the best option for the OWS case (Figure 8), does not seem preferable also for CO$_2$RR or NRR. In fact, Figure 9 does not evidence a clear distinction between the performances of PEC and PV/EC technologies. Also in this case, testing essentially only at room temperature, and without the use of sun-concentrations has been reported.

In principle, maximizing the overpotential difference between that necessary for CO$_2$RR or NRR reactions, and that for the side HER reaction (which should be the highest possible) would allow to operate at high $J$, while maintaining high the Faradaic selectivity (typically indicated also as Faradaic efficiency), at least with respect to the side HER reaction. However, this is not a trivial task because it is necessary to activate hydrogen to allow the hydrogenation of the adsorbed CO$_2$ (or the products of further reduction) or N$_2$ while avoiding the recombination between activated hydrogen species that would form H$_2$. Thus, an optimal compromise is difficult to obtain.

Moreover, HER is a "simple" reaction since it involves only two electrons (eq. 6), while CO$_2$RR and NRR usually require a higher number of electrons, except for CO or HCOOH production (eq.s 8a and 8b). Faradaic efficiencies for these reactions above 90% have been achieved with high $J$ in electrochemical reactions. For example, Wu et al.[206] reported for a PbO$_2$ electrode in the electroreduction of CO$_2$ to HCOOH in an ionic liquid electrolyte a Faradaic efficiency of 95.5% and a current density of 40.8 mA·cm$^{-2}$ simultaneously. However, selectivity decreases when products of CO$_2$ reduction, involving >2 electrons, are targeted such as CH$_3$OH (a 6e$^-$ reduction, eq. 9) or products involving also C-C bond formation (eq.s 10-12 as an example). In addition, the surface coverage by adspecies also plays a crucial role, and in fact by increasing the pressure of operation in CO$_2$RR, the Faradaic selectivity can be improved.[207] The lower surface coverage in N$_2$ with respect to CO$_2$, due to strong N≡N bond, explains why the Faradaic selectivity is typically lower in NRR with respect to CO$_2$RR.[87,88]

Note also that reliable data on carbon or nitrogen balance in CO$_2$RR and NRR are usually not given. Significant crossover from the cathodic zone (where CO$_2$RR and NRR occurs) to the anodic zone of the cell may be present (or viceversa), but often this is not an aspect considered. Analyzing the experimental $J$ with respect to the $J$ calculated from the current density needed to form all the detected products is a straight manner to have indication on these aspects as well as on other effects (for example, other side reactions including the reduction of the electrocatalyst). It should be thus part of the protocol of experimentation, although typically not made.

As final comment, it may be argued that an alternative strategy, giving the better performances in OWS than CO$_2$RR

**Table 3** Comparison of representative literature results concerning CO₂RR. GDE: gas diffusion electrode.

| Cell Design | Components | $J_{SC}$ / $J_{op}$ (mA·cm⁻²) | Product /FE (%) | $\eta_{STF}$ (%) | Ref |
|---|---|---|---|---|---|
| PV/EC | PV: GaInP/GaInAs/Ge EC: CuO/SnO₂ | $J_{op}$ = 11.57 | CO/~80 | 13.4 | 189 |
| PV/EC | PV: Si EC: Se-(NiCo)Sₓ/ (OH)ₓ \| Ni foam; Cu \| GDE | $J_{SC}$ ~ 55 $J_{op}$ = 52.3 | C₂H₄/35 EtOH/32 | 3.9 | 206 |
| PV/EC | PV: Si EC: IrO₂; Cu₂O | $J_{SC}$ - $J_{op}$ = 20.3 | C₂H₄/32 C₂H₅OH/10 | 1.5 0.45 | 207 |
| PV/EC | PV: perovskite EC: CuO; CuO | $J_{SC}$ = 27.2 $J_{op}$ = 18 | C₂H₄/34 | 2.3 | 208 |
| PV/EC | PV: GaInP/GaInAs/Ge EC: Ni foil; Ag\|GDE | $J_{SC}$ = 14.6 $J_{op}$ = 14.4 | CO/99 | 19.1 | 209 |
| PV/EC | Cu(In,Ga₋₁) (SᵧSeᵧ₋₁)₂//Au | $J_{SC}$ = 3.5 | CO/91 | 4.2 | 210 |
| PV/EC | c-Si PV// Ag/IrOₓ// MWCNTs/RuCP | $J_{SC}$=6.5 | HCOOH/>80 | 7.2 | 213 |
| PV/EC | PV: Perovskite (3.1 V) EC: Anodised Au // IrO₂ | - | CO/~90 | 6.5 | 190 |
| PV/EC | PV: Si (2.2 V) EC: Mn complex // FeOₓ/NiOₓ | $J_{SC}$ = 1 | CO/~80 | 3.4 | 191 |
| PV/EC | PV: Si (2.7 V) EC: Cu–Zn // Ni | $J_{SC}$ = 5 | CO/<20 | 4.3 | 192 |
| PV/EC | PV: Si (2.7 V) EC: Porous Bi // IrO₂ | $J_{SC}$ = 10 | HCOOH/~95 | 8.5 | 214 |
| PV/EC | WSe₂/PV-a-si-3jn // CoOₓ/PV-a-si-3jn (tandem) | - | CO/~24 | 4.6 | 196 |
| PV/EC | Pd/C/Ti // Ni/TiO₂/InGaP/GaAs | 8.7 | HCOOH/~94 | ~10 | 193 |
| PEC | Si//Ag/Cu | ~ 30 | C₂H₄,C₂H₅OH, C₃H₈O/80 | - | 194 |
| PEC | GaInN/Si//Cu | 1.7 | C₂H₄//17.5 | 0.41 | 211 |
| PEC | BiVO₄/FeOOH //NiOOH | 0.24 | CO/45 | 0.29 | 212 |
| PEC | CuFeO₂/CuO // Pt | <0.2 | CO/~90 | ~1 | 195 |
| PEC | Ru complex/InP // Reduced SrTiO₃ | 0.1 | HCOOH/~71 | 0.14 | 215 |
| PEC | Ru-complex/InP // TiO₂ | - | HCOOH/~70 | 0.04 | 216 |
| PEC | Ru-complex/TiO₂/Fe₂O₃/ Cr₂O₃ // TiO₂ | ~0.1 | HCOOH/~79 | 0.,15 | 198 |
| PEC | Co-complex/perovskite // CoOₓ/BiVO₄ | ~1 | CO/~25 | 0.08 | 217 |
| PEC | ZnO–ZnTe@CdTe-Au // Co-HCO₃/Ni foam (assisted perovskite MAPI) | 0.85 | CO/~80 | 0.43 | 197 |
| PEC | Bio/TiN // FeOₓ/BiVO₄ (assisted perovskite) | <1 | HCOOH/~79 | 0.08 | 199 |
| PEC | RuRe-complex/CuGaO₂ // CoOₓ/TaON | <0.05 | CO/~41 | - | 218 |

and NRR, is to produce H₂ by PV/EC approach, and then use H₂ in catalytic (thermal) processes of CO₂ or N₂ conversion. This solution, however, is not efficient in terms of integration in small-scale devices, and thus does not adapt to the development of artificial-leaf type (distributed) devices.

Direct CO₂RR and NRR in PEC (or analogous) devices remain a preferable strategy for several reasons: (i) the overpotentials necessary in H₂ formation (HER) and in use H₂ (need to activate and compress H₂) are eliminated, due to the direct use of protons/electrons in N₂ or CO₂ reduction, (ii) mild operative conditions are used, and (iii) costs are potentially lower using an integrated device rather than two separate units (electrolyser and catalytic unit). If improved electrocatalysts could be developed, a potential reduction of both fixed and operative costs can be achieved.

## The case of CO₂

Galan-Mascaros,[11] Creissen and Fontecave[155] recently reviewed the best performing results about PV/EC and PEC systems for CO₂RR commenting the low number of examples of unbiased solar CO₂ reduction processes. Table 3 reports the comparisons of $\eta_{STF}$ and $J$ of representative literature examples for CO₂RR.

Cheng et al.[209] reported top performances for PV/EC approach of a $\eta_{STF}$ up to 19% (at 1 SUN) with current density of about 14 mA·cm⁻² but only towards CO formation and using a costly triple-junction solar cell. The proof of the formation of only CO (and carbon balance) is not fully convincing. The authors claim long-term stability. Outdoor operations allowed a peak solar-to-CO efficiency of 18.7%, stable for >150 h and $J$~15 mA·cm⁻². Schreier et al.[189] reported a $\eta_{STF}$ of about 13%, also to only CO (but slightly lower Faradaic selectivity) and with a slight lower current density (about 14 mA·cm⁻²). They also used a GaInP/GaInAs/Ge costly triple-junction solar cell, but earth-abundant materials for the electrocatalysts, differently from Cheng et al.[209]

Zho et al.[193] reported a $\eta_{STF}$ of about 10% to formate, with current density of about 9 mA·cm⁻² also using a costly InGaP/GaAs PV component. A $\eta_{STF}$ of about 8.5% to formate with current density about 10 mA·cm⁻² was reported by Pia et al.[214] also using a triple-junction PV cell, but based on less-costly Si. However, a costly component such as IrO₂ was used as electrocatalyst. Kato et al.[213] reported a $\eta_{STF}$ ~7% to formate using a Si-based PV cell and costly noble-metal based electrodes. The cell consists of five stacked electrodes (electrically parallel connected) and six series-connected single-crystalline Si PV cells (area ~1,000 cm²). The electrodes contain quite expensive elements (anodes based on IrO₂ and cathodes on Ru complex polymer) whose stability is unclear. The cell is without a membrane and operates at 1.85 V and 6.30 A.

Lower $\eta_{STF}$ values have been reported in producing SFs involving a higher number of electrons for the reduction than the 2e⁻ for CO and HCOOH. For example, ethylene was obtained at $\eta_{STF}$ = 2.9% combining Si solar panels with an electrolyser.[207] Membrane-less electrolysers powered by Si PV cells reached $\eta_{STF}$ >3% to CO[191] using earth-abundant catalysts with $J$~1 mA·cm⁻¹.

The integrated PEC approach for CO₂RR results in lower efficiencies, with $\eta_{STF}$ typically below 1% (Table 3).[11] Zhou et al.[219] reported $\eta_{STF}$>10%, but for a PEC device incorporating a photoanode made of a TiO₂/InGaP/GaAs heterostructure decorated with a Ni OER catalyst which drives a dark Pd/C/Ti cathode producing formic acid at >94% Faradaic efficiency.

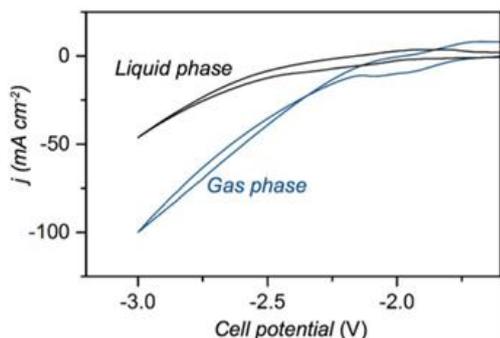

**Figure 10.** Comparison of the cyclic voltammograms measured on liquid- (black line) and gas- (blue line) based flow cells. Reproduced from ref.[226] Copyright American Chemical Society @2017.

Current density is about 8.7 mA·cm⁻². Incorporation of Si-heterojunctions in the photoanode and photocathode yielded $\eta_{STF}$ > 4% with CO as major product using ionic liquids as electrolyte and a WSe₂-based cathode.[196] A TiO₂/p-Si photocathode modified by cobalt bis(terpyridine) for CO₂ to CO production,[220] but $J$ were quite low (about 0.1 mA·cm⁻²). Data for $J$ were not always reported directly, so they could be deduced from the other results, but indications in Table 3 evidence that typically the values are <1 mA·cm⁻².

Few results have been reported for integrated PEC devices in producing C2+ products,[194] although a recent review discussed the photocatalytic CO₂ reduction to C2+ products.[221] Combined with an external PV cell, an electrolyser based on a Cu/Ag/TiO₂ photocathode results in 60% Faradaic efficiency to form a mixture of C2/C3 products. A tandem device was developed by coupling a Si photocathode with two series-connected semi-transparent CH₃NH₃PbI₃ perovskite solar cells, achieving an efficiency for the conversion of sunlight to hydrocarbons and oxygenates of 1.5% (3.5% for all products, which include H₂, CO - the dominant products -, ethylene, ethane, formate, propionaldehyde, allyl alcohol, ethanol, 1-propanol, and other C2+ products).

Values of $J$ range from 2 to 3 mA·cm⁻² depending on the electrolyte concentration. Table 3 evidences two other aspects: (i) focus is on two electrons reductions (CO, HCOOH), while more complex reactions will be likely preferable; one case,[194] however, reports good performances to C2/C3 products, although producing a range of different chemicals, and (ii) $J$ are generally low, and only slightly better when a PV component is integrated in the system, although typically with problems of stability, mainly related to photocorrosion.

The CO₂ electrocatalytic reduction (thus without direct coupling with a photoanode or a PV cell) is an area of extensive current studies and several works were reported about the electroreduction of CO₂ at elevated $J$ (in the range of hundreds mA·cm⁻²).[222,223] Such results have been obtained by exploiting device architecture that overcome the solubility problem of CO₂ into water-based electrolyte. For example, gas-phase flow cells have been used for the production of alcohols (methanol, ethanol and n-propanol) and CO at, respectively, $J$ of 180 and 60 mA·cm⁻² and FE~40 (cumulative) and >90%.[224,225] Furthermore,

Li et al.[226] demonstrated that gas-fed flow cells deliver up to twice more $J$ (reaching also 200 mA·cm⁻² with a FE ~50% towards CO) with respect to those based on liquid electrolyte (as shown in Figure 10).

Weekes et al.[222] and Endrődi et al.[223] discussed the role of cell architecture in the electroreduction of CO₂, while more detailed aspects are out of the scope as well as the analysis of the recent development in electrocatalytic CO₂ reduction, which for formate production from CO₂ reached recently extremely high current densities (~930 mA·cm⁻²) with high Faradaic efficiency (FE$_{HCOO-}$ = 93%) on defective Bi₂O₂CO₃ nanosheets.[227]

While thus high $J$ and FE are possible in the electrocatalytic approach (with an external source of potential and electrical current), the direct integration of a photoactive element in the device (photoanode, PV cell) allows largely lower $J$ values, although with other advantages as earlier discussed.

Based on the presented results, the PV/EC design can be the best solution for reaching high-$J$ also for CO₂RR, but the EC component must be properly engineered to account for the mass transport limitations that influence this cell performances. Cheng et al.[209] (Table 3) reported an example of such synergistic approach using a triple junction solar cell (GaInP/GaInAs/Ge) to power an Ag-based GDE. The coupling between the photoactive element and the devices is optimal ($J_{OP}$~$J_{SC}$) and the electrocatalyst shows high selective FE (99%). This result was obtained by optimizing many parameters (electrolyte pH and compositions) in relation to $J$. Another work, by Huan et al.[208], used a perovskite-based solar cell and a CuO-based EC to reduce CO₂. Although the performances are modest (as shown in Table 3), the use of a perovskite solar cell is an interesting path towards low-cost PV components for PV/EC systems.

However, as pointed out by Burdyny and Smith,[156] current studied on CO₂RR mainly focus on low-$J$ conditions, thus the information obtained within this framework may not be useful when considering systems delivering high-$J$. In particular, it is not still clear if CO₂RR proceeds as a three- or two-phases reaction, moreover some electrolytes (used for high selectivity) can exhibit large ohmic losses when operating at high-$J$ conditions, which hinders their use in practical industrial-scale devices.

### Nitrogen reduction reaction

Reducing N₂ to NH₃ is a challenging and intriguing chemical reaction since the N₂ molecule is rather inert due to the high stability of the N≡N triple bond. In contrast to many papers on electrochemical/catalytic approaches for NRR,[87,88,228–231] few reports can be found on this reaction with PEC devices. The most interesting examples bare in common the use of plasmonic nanoparticles:

- Au nanoparticles on B-doped etched Si wafer (to form a pillared-type nanostructure; a Cr layer with thickness ~50 nm is sputtered on the top of the substrate to protect from corrosion, but Cr acts also as sacrificial agent) giving an ammonia yield ranging from about 22 to 35 mg$_{N2}$·m⁻² on increasing light intensity from 1 to 7 SUN; $\eta_{STF}$ and $J$ are not given, but estimated to be <0.1% and <0.1 mA·cm⁻²,

respectively)[232]

- Ag nanoparticles on black silicon photocathodes[233] giving a current density (at the maximum $NH_3$ production rate of 2.87 $\mu mol \cdot h^{-1} \cdot cm^{-2}$) at −0.2 V, thus bias assisted) of about 0.12 $mA \cdot cm^{-2}$ with $FE_{NH_3}$ ~41% (deactivation is observed in about 2h of experiments); $\eta_{STF}$ is not given, but estimated to be <0.1%

- Au-poly (tetrafluoro ethylene)-Si photocathodes taking advantage of a hydrophobic environment to preclude HER[234]; ammonia yield at the best (-0.2 V, thus bias assisted) is about 19 $\mu mol \cdot h^{-1} \cdot cm^{-2}$ with $FE_{NH_3}$ about 39%; $\eta_{STF}$ is not given, but estimated to be <0.1%; for J is nearly zero without applied potential and increases by applying a negative potential, but still remaining below ~1 $mA \cdot cm^{-2}$

- Au-SrTiO$_3$ plasmonic photoanodes, associated with a Zr cathode[235]; without applied bias the $FE_{NH_3}$ is nearly zero and J about 2 $\mu A \cdot cm^{-2}$ with an IPCE at the best of 0.07% at 600 nm (thus much lower $\eta_{STF}$).

Therefore, current state-on-the-art in NRR in PEC devices indicates extremely low J (< 0.1 $mA \cdot cm^{-2}$) and $\eta_{STF}$ (estimated significantly below 0.1%), and the typical need of applying an external bias to obtain measurable results.

In some case, the term photoelectrochemical was used to indicate an electrocatalytic cell, to which an external bias was mandatory for its function. For example, Mushtaq et al.[236] claimed to have investigated the PEC $N_2$ reduction to $NH_3$ in a two-compartment H-type cell separated by the Nafion membrane. However, there is no photoanode or external PV cell, and it is thus an electrochemical (or electrocatalytic, EC) cell, not a PEC device. Similarly, Zheng et al.[237] in reviewing the photoelectrochemical conversion of $CO_2$ and $N_2$ into useful products on $TiO_2$ (titanate) based nanomaterials, indicated a series of studies of PEC NRR.[235,238–242] However, the results were instead referring to different situations. Xu et al.[238] used a 3D-printed hierarchical porous $TiO_2$ scaffold immersed in a solution and illuminated, thus like PP case, but no values of J or $\eta_{STF}$ were reported.

Shi et al.[239] indicated photoelectrochemical measurements but measuring the photocurrent of an Au/$TiO_2$ electrode applying +0.3 V potential and the observed photocurrent was very low (< 0.01 $mA \cdot cm^{-2}$). Oshikiri et al.[235] used Au-doped SrTiO$_3$ photoelectrode with a two-compartment cell, but separated by the NbSrTiO$_3$ dense layer, illuminated by a Xe lamp. The closure of the electronic circuit is unclear, and a sacrificial agent (ethanol) is present in the anodic chamber. Li et al.[240] used plasmon-enhanced Au-$TiO_2$ (nanorod array) photoelectrode immersed in an aqueous solution, which is a PP-like approach, rather than a photoelectrochemical approach as reported and the resulting $\eta_{STF}$ is below 0.002%. Ye et al.[241] also indicated a photoelectrochemical approach based on nanojunctions assembled from $MoS_2$ nanosheets and $TiO_2$. They used a two-chamber H-type cell divided by a Nafion 211 membrane, with the $MoS_2@TiO_2$ (deposited on carbon substrate) on the illuminated side. However, to observe $NH_3$ formation it was necessary to apply an external bias and the results were unclear in separating the contribution of illumination from that of the electrochemical cell.

Li et al.[242] recently reported PEC NRR using a $Mo_2C/C$ heterostructure, and $BiVO_4$ as photoanode. Also here an external bias (0.2 V) was applied which doubles the production of $NH_3$. No indication is given in terms of $\eta_{STF}$ and the estimated specific current density to ammonia was below ~0.03 $mA \cdot cm^{-2}$ in the absence of an external bias and thus low.

Wang et al.[243] also reported the photoelectrochemical $NH_3$ synthesis using a Co-phosphate/Ti-Fe$_2O_3$ photoanode and Co-SAC (single-atom catalyst) cathode. However, the PEC device has no separate compartments. In addition, an external bias should be applied to observe both photocurrent (which is enhanced with respect to case without illumination) and formation of $NH_3$. Thus, it is a light enhanced electrochemical NRR, rather than a PEC NRR as indicated.

As emerges from this analysis, data on the PEC approach for NRR are limited, and results without an external bias, or the use of sacrificial agents, are not available. Moreover, both J and $\eta_{STF}$ are either not given or extremely low. A question is thus why these poor performances not only with respect to OWS, but also to $CO_2$RR.

One of the differences between $CO_2$ and $N_2$ reduction cases is related to the coordination of the molecules on the surface of the electrocatalyst. Both the strength of C=O and N≡N bonds is different, as well as the availability of $d$ orbitals for coordination, making more challenging $N_2$ molecule activation. In Fe or Ru catalysts for heterogeneous (thermal) catalysis, dissociative chemisorption of $N_2$ is the first and rate limiting step. The general agreement is that in electrocatalysis, hydrogenation of undissociated coordinated $N_2$ molecules is the rate determining step.[87] End-on $N_2$ molecule coordination at mono- or bi-metallic sites is the first step of $N_2$ coordination, followed by sequential or simultaneous multi proton-assisted electron transfer.

Note that, differently from conventional electrochemical electron transfer, this process is typically considered to occur on chemisorbed $CO_2$ or $N_2$ molecules, rather than through an outer shell electron transfer mechanism as in electrochemistry theory.

The difference in $CO_2$ versus $N_2$ chemisorption on the electrocatalyst is a main reason explaining the different behaviour, both in terms of activity (and J) and Faradaic efficiency (lower surface coverage by $N_2$ would imply also easier side HER).

In addition, a further limiting factor for J is $N_2$ solubility, which is also related to NRR selectivity because a higher selectivity increases the surface coverage by $N_2$.[78] Note, however, that as commented for $CO_2$ and outlined in Figure 5, it is possible to operate with GDL-type electrodes and cell configurations, which largely overcome the problem of $N_2$ solubility (as assessed previously for $CO_2$RR). This aspect, however, is rarely recognized.[81] Thus, the design of devices should be focused at maximizing sites coordinating molecular $N_2$ in a form susceptible for proton-assisted electron transfer. On the other hand, cell design, which can force the creation of an enhanced local pressure of $N_2$ (or enhanced chemisorption) at the electrocatalyst surface would be equally relevant. Creating a suitable overlayer on the catalyst which enhances the

Perspective

N₂ adsorption and create a pseudo-virtual higher $N_2$ pressure at the electrocatalyst is a way to explore, but not attempted as far as we know.

An improved design of the electrocatalysts is necessary for an efficient (photo) electrochemical $NH_3$ synthesis, since current results are still far from those required in industrial applications (for both FE and $J$ values).[88] While undoubtedly there is the need to a better mechanistic understanding of the reaction and the identification of the active sites, current approaches (including theoretical methods) are largely unable to properly identify these aspects. In front of an exceptionally large range of different indications on the active sites and reaction mechanisms, all the results fall within a limited range, pointing out that the aspects identified are likely not those relevant.[87,88] Arguably, a better analysis of the parallelism between enzymatic (*Nitrogenase* cofactor) and electrocatalytic mechanism could be a great source of inspiration in the design of novel electrocatalysts. It is worth noting that the *Nitrogenase* enzyme is also able to convert $CO_2$ to C2+ products, indicating that an analysis of electrocatalytic systems with active sites able to perform both reactions, could be an interesting route to design novel improved electrocatalysts. However, also this aspect has been scarcely recognized in the literature.[87]

Among the best performing non-precious-metal electrocatalysts for $NH_3$ synthesis is a Bi-based material which shows a FE of 66% at $J$ = 4.2 mA·cm⁻².[244] However, doubt exists on the results, and for example Choi et al.[245] proved that Bi does not exhibit measurable electrocatalytic activity for NRR. Bi nanocrystal catalysts show for this reaction a fast deactivation and since these are prepared starting from Bi-nitrate, the detected ammonia could derive reasonably from the residual nitrate species present: Bi is not a conductive material and Bi nanocrystals in the 50-100 nm range cannot be effective electrocatalysts. This exemplifies the care necessary in analysing literature data, since NRR still is not an area in which well consolidated results are present.

In terms of target performances (for the electrocatalytic reactions), McPherson et al.[246] indicated a FE >90% at $J$>300 mA·cm⁻², that agrees well also with our findings.[88] However, such $J$, as previously shown, cannot be reached in a PEC device, and thus the only possible approach is PV/EC. However, this approach is effective for more simple reactions (two electrons), but is questionable, and never demonstrated up to now, for multielectron reactions such as $NH_3$ synthesis (6 electron reduction of $N_2$) which, in addition, competes with a two-electron reduction reaction (formation of $H_2$).

The production of $NH_3$ from $N_2$ reduction through the PEC architecture would thus be at this stage the most promising approach, although clearly $J$ (and thus productivities) are lower than those indicated before as target. The lower productivity is compensated, however, to the possible cost reduction in having a single integrated device able to directly use sunlight to convert $N_2$ and $H_2O$ to $NH_3$. As commented for $CO_2RR$, the use of sun-concentration is likely necessary also for NRR, but in addition critical is the identification of solutions to enhance the selective coverage by $N_2$ of the electrode, for the motivations discussed before. A solution could be a different approach, where in the anodic part $N_2$ is oxidized to $NO_x$ and then $NO_x$ are reduced in the cathodic part, being their reduction faster and chemisorption higher with respect to the $N_2$ case. Wu et al.[247], for example, showed that nitrate reduction to $NH_3$ has a larger Faradaic efficiency (~75%) and yield rate (~20.000 μg·h⁻¹·mg$_{cat}$⁻¹) with respect to $N_2$ reduction to $NH_3$. However, it is necessary to provide the $H^+/e^-$ for the reduction of $NO_x$ and thus cell design could be critical. No attempts in this direction have been reported as far as we known, but it is a valuable direction to explore to overcome the still critical limitations present in NRR electrocatalytic approaches (including with an integrated photoactive element).

Note, in addition, that from a practical perspective, a major issue would be the need to separate $N_2$ from air. A device which can be fed directly with air (rather than first separate $N_2$ and $O_2$ in pure streams) would be necessary, or at least develop NRR electrocatalysts/devices with limited sensitivity to $O_2$ contaminations in the $N_2$ feed (current results are all obtained with ultrapure $N_2$ flow). To avoid competition of $O_2$ in $N_2$ reduction, the electrocatalyst should integrate a $N_2$-permeoselective membrane. This is a relevant future direction of R&D, but to our best knowledge not yet investigated.

As a general comment, the $N_2$ reduction potential is slightly more positive than the reduction potential of $H_2O$.[248] Thus, in principle, materials used for $H_2O$ reduction should work also for NRR, but only when specific sites able to activate selectively $N_2$ are present.

For example, some attempt has been also made to use an efficient photocathode for HER, decorated with cocatalysts that are active towards $N_2$ activation. Within this idea, CuO has been used as photocatalysts for $NH_3$ synthesis, resulting in good FE (~20% for $NH_3$), although low $J$ were reported (~0.1 mA·cm⁻²).[248] In another study, a $BiVO_4$ photoelectrode, decorated with $MnCO_3$ and C has been shown to have good current densities ($J$ ~ 2 mA·cm⁻²), about 8-times higher than pristine $BiVO_4$. However, no data were reported for FE, and thus $J$ was not specific in ammonia formation. Stability, in addition, was limited.[249]

Several studies have been also focused on the direct photocatalytic NRR.[83,96,250–252] As commented in these reviews, photocatalytic NRR faces many obstacles among which the low yield of ammonia, low quantum efficiency, fast recombination of electrons and holes, and the possibility that nitrogen may be further oxidized to nitrite or nitrate.

Finally, note that there is an ample debate about the reliability of NRR results reported in literature, particularly for the case of electrochemical synthesis. Among the possible sources of errors are the (i) contamination by N species of the electrocatalyst or other materials used (membrane, etc.) and thus that the observed $NH_3$ derives from this contaminating N rather than from gaseous $N_2$, and (ii) presence of N-species ($NH_3$ itself or NOx which can be then reduced to $NH_3$) in the $N_2$ feed to the electrocatalytic reactor. Tests with labelled $^{15}N_2$ are typically required to validate that $NH_3$ derives from $N_2$. The formation of other products, hydrazine in particular, and the crossover of ammonia through the membrane are other elements of possible error.

## Conclusions and perspectives

The production of SFs ($H_2$ from water and products of $CO_2$ and $N_2$ reduction) is a promising solution towards the foundation of a green and sustainable economy. Several technologies have been proposed with the aim to optimise the performances of the resulting devices such as particulate photocatalysis (PP), photoelectrochemical (PEC) catalysis and photovoltaic driven electrocatalysis (PV/EC). In this perspective, we analysed the role of current density ($J$) as a fundamental figure of merit because i) it allows a more direct comparison between different approaches for SFs production and ii) it gives valuable information about the physico-chemical processes involved during the device operations. The latter aspect is crucial since a thorough knowledge of the working mechanisms (and associated limits) is fundamental to progress in the developments of technologies for SFs production. For convenience, the physico-chemical processes behind the conversion of sunlight into chemical fuels were divided in four steps: i) photon absorption and generation of free charge carriers, ii) separation and transport of charge carriers, iii) catalytic conversion and iv) mass transfer. Careful analysis of $J$ gives information about each step, in particular:

- The short circuit current density ($J_{SC}$) is the maximum current that can be photo-generated by the light harvester used in the device. Many devices use one single light-harvester, thus the maximum theoretical $J_{SC}$ is 73 mA·cm$^{-2}$ for a material with zero energy bandgap and one SUN irradiation. This has two main consequences i) low band-gap materials are needed to achieve high $J$ (but this requirement contrasts that of minimum photovoltage needed to trigger the desired reaction) and ii) $J$ comparable to those used in existing technologies for chemical fuels production (~0.5-2 A·cm$^{-2}$) cannot be practically achieved. When values of $J_{SC}$ are lower with respect to the theoretical ones, the production of free electron/hole pairs is hindered by other processes such as exciton formation, that can be mitigated by proper engineering of the dielectric properties of the light-harvester (which influence the exciton binding energy).

- Several mechanisms can be responsible for $J$ values lower than $J_{SC}$ (at applied biases ≠ 0V), arising from the other three steps herein discussed. The separation and collection of the photo-generated charge carriers can suffer from recombination phenomena: monomolecular, bimolecular and Auger processes are the main ones. Thus, a careful assessment of the recombination constants is fundamental to set realistic goals for the maximum $J$ obtained by a device assembly. The probability of occurrence of such phenomena is further increased by a low catalytic activity of the cocatalysts: if charge carriers are not quickly used to trigger the reaction of interest, they have a greater chance to recombine. Finally, sluggish motion of the ions in the electrolyte hinders the efficient "refuel" of the reactants adsorbed on the cocatalyst surface, which slows down charge carrier motion and so $J$. As a result, all these steps must be properly accounted for to maximise the performances of the device.

Each technology has its pros and cons to produce a particular SF. For example, with $H_2$, the most promising architecture is the PV/EC which can deliver remarkably high $J$ thanks to the proper series/parallel connection of several PV units. Large scale demonstrations of such systems have been studied, with very encouraging results. However, the costs associated to both the PV and EC components have questioned the economic impact that such technology would pose on the market. PEC approach still is a hot topic among the scientific community (for converting small molecules, i.e. $H_2O$, $CO_2$, and $N_2$)[245] since in this case both charge carrier generation and catalytic processes are carried out by a single component. Here, the produced $J$ would be limited by the absorption properties of the light harvester and in general it cannot exceed 73 mA·cm$^{-2}$ so, there is a trade-off between architecture complexity/cost and performances.

Differently, for the case of products of $CO_2$ and $N_2$ reduction, due to various additional issues (with respect to OWS) making these reactions ($CO_2$RR and NRR) more challenging and still far from the application. As commented before in analysing the state-of-the-art, PEC approach appears the more promising for these reactions, but $J$ values are low. There are scattered results and not systematic regarding how to improve cell design and $J$. There is the need to intensity R&D on these aspects, but it requires to first go back in understanding better the crucial parameters determining the performances and how to design an optimal cell to minimize the current limitations.

A general conclusion which can be derived from the discussion presented here is that from an implementation perspective of producing solar fuels from $CO_2$ and $N_2$ conversion in PEC or PV/EC devices, the need to operate these cells in conjunction with sun concentration (in the 50-100 SUN range) clearly emerges. However, this is not just an additional parameter to investigate, because it requires a fully redesign of both materials and cell to operate efficiently and stable under these conditions. Most of the systems and design approaches development up to now have severe limitations to operate under these conditions, and thus a general comment emerging from the discussion is the need to rethink current approaches to produce solar fuels. For NRR, there is also the additional requirement of developing efficient solution increasing the surface coverage of the electrocatalyst by $N_2$. The novel possibility to use PEC devices to oxidize $N_2$ to $NO_x$ on the anodic side and then reduce to $NH_3$ the $NO_x$ on the cathodic side has been shortly introduced, although is a novel unexplored direction.

In conclusion, this perspective paper provides indications and clues to identify relevant directions to explore to implement solar fuels distributed production in PEC or PV/EC type of devices. The concept of current density is a useful figure of merit to address these aspects.

## Acronyms

CB       conduction band
CBM      conduction band minimum
CES      chemical energy storage

CO₂RR  CO₂ reduction reaction
EC  electrocatalytic
FoMs  figure of merits
GDL  gas-diffusion layer
HER  hydrogen evolution reaction
IPCE  incident-photon-to-current-collected-efficiency
LO  longitudinal optic (phonon)
NRR  nitrogen reduction reaction
OER  oxygen evolution reaction
OWS  overall water splitting
PA  photoanode
PC  photocathode
PEC  photoelectrocatalytic (or photoelectrochemical)
PV/EC  photovoltaic-driven-electrocatalysis
PP  particulate photocatalysis
PV  photovoltaic
RHE  reversible hydrogen electrode
SF  solar fuel
STF  solar-to-fuel
STH  solar-to-hydrogen
VB  valence band
VBM  valence band maximum

## Conflicts of interest

There are no conflicts to declare.

## Acknowledgements

This work was made in the frame of the ERC Synergy SCOPE (project 810182), EU DECADE project (nr. 862030) and PRIN 2017 project MULTI-e nr. 20179337R7 and CO₂ ONLY project nr. 2017WR2LRS, which are gratefully acknowledged. The authors also thank the initiative SUNERGY (www.SUNergy-initiative.eu) which is dedicated to promoting solar fuels to substitute the use of fossil fuels and unlock the renewable energy future.